\documentclass[prl,twocolumn,showpacs,preprintnumbers,superscriptaddress]{revtex4}

\usepackage{times}
\usepackage{subfig}
\usepackage{bm}
\usepackage{graphicx}
\usepackage{amsbsy}
\usepackage{amsmath}
\usepackage{amsfonts}
\usepackage{amsthm}
\usepackage{float}
\usepackage{color}

\begin{document}

\theoremstyle{plain}
\newtheorem{theorem}{Theorem}
\newtheorem{lemma}[theorem]{Lemma}
\newtheorem{corollary}[theorem]{Corollary}
\newtheorem{proposition}[theorem]{Proposition}
\newtheorem{conjecture}[theorem]{Conjecture}

\theoremstyle{definition}
\newtheorem{definition}[theorem]{Definition}

\title{A study of Quantum Correlations in Open Quantum Systems}
\author { Indranil Chakrabarty}
\email{indranil@iopb.res.in}
\affiliation{Institute of Physics, Sainik School Post, 
Bhubaneswar-751005, Orissa, India}
\author{Subhashish Banerjee}
\email{subhashish@cmi.ac.in}
\affiliation{Chennai Mathematical Institute, Padur PO, Siruseri- 603103, India}
\author{Nana Siddharth}
\email{nanasid@cmi.ac.in}
\affiliation{Chennai Mathematical Institute, Padur PO, Siruseri- 603103, India}
\begin{abstract}
In this work, we study quantum correlations in mixed states. The states studied are modeled by
a   two-qubit system interacting with its environment
via a quantum non demolition (purely dephasing) as well as dissipative type of  interaction. 
The entanglement dynamics of this two qubit system is analyzed. We make a comparative study  of various measures of quantum correlations, like Concurrence,  Bell's inequality,  Discord and Teleportation fidelity, on these states, generated by the above evolutions. We classify these evoluted states on basis of various dynamical parameters like bath squeezing parameter $r$, inter-qubit spacing $r_{12}$, temperature $T$ and time of system-bath evolution $t$. In this study, in addition we report the existence of entangled states which do not violate Bell's inequality,  but can  still  be useful as a potential resource for teleportation.  Moreover we study the dynamics of quantum as well as classical correlation in presence of dissipative coherence.
 
\end{abstract}
 

\maketitle

\section{Introduction}
Entanglement lies at the heart of quantum mechanics.  In the last few years a lot of research has been done in the field of detection and quantification of 
entanglement. The original  method for  detection of entanglement was  Bell's Inequality \cite{bell,chs}.  
Bell's inequality is  derived from  assumptions of a general, local hidden-variable theory (LHVT). 
This inequality is unique in the sense that all local hidden variable models  satisfy it. 
In other words, the inequality  derived by Bell  has clearly demonstrated that some quantum mechanical predictions, 
in the ideal scenario, cannot be reproduced by any LHVT. 
Not only that,  the violation of Bell's inequality is a strong signature for the inseparability of a quantum state. 
However,  Bell's inequality is unable to detect all possible entangled states as there are examples of entangled states which does not violate it \cite{wer}.
Thus one can conclude that  violation of Bell's inequality implies entanglement, 
but the presence of entanglement does not necessarily imply  violation of the inequality.

There is thus a need for a measure that
can quantify the amount of entanglement present in the system.  Out of many measures a very 
convenient one is  concurrence \cite{woo}.
Apparently,  it may look that there is no  difference in quantifying  inseparability and   quantumness. However, in general,
it is not so.  Even though it is quite well
established that entanglement is essential for certain kinds of quantum-information tasks like teleportation and super-dense coding,  
the precise role of entanglement in quantum
information processing still remains  an open question. 
It is not clear, whether  all information-processing tasks that can
be done more efficiently with a quantum system than with a comparable classical
system require entanglement as a resource. Indeed, there are  instances where
it is possible to do more with quantum rather than the classical, 
in the absence or near absence of entanglement \cite{braun,ken,bih,mey,kni}.
This tells us that one cannot ascribe the mere absence of entanglement as  a signature of classicality.  
As an example, we can talk about the
DQC1 (deterministic quantum computation with one bit) model where the system has very little entanglement,  and even that little vanishes asymptotically,
yet it provides an exponential speedup \cite{kni,ani}.

This raises an important question that if not
 entanglement  what explains all the quantum advantages? 
 Most likely, it has something to do with the structure and inherent non-locality
of quantum mechanics. Interestingly, it is known that non-locality and entanglement
are not equivalent features.  Entanglement is the feature that having complete
information about the subsystems does not reproduce the complete information
of the whole system. 
Not surprisingly,  this is not the only characterization of a quantum system.
For instance, the collapse of one part of a subsystem after
measurement of another is another feature which is unique to quantum systems and does 
not have any classical analogue. The quantity that tries to capture this unique feature is the
quantum discord \cite{oll,hen,luo}.

Quantum discord tries to quantify all types of quantum correlations 
including entanglement.  It must be emphasized here that  discord supplements the measures
of entanglement that can be defined on the system of interest. It aims to capture
all the non classical correlations present in a system,  including
entanglement. Other measures of quantum correlation, similar to discord,
are the quantum dissonance \cite{mw10},  that aims to capture quantum correlations
in separable states and measurement induced disturbance \cite{luo08}.  

It may appear  that analyzing Bell's inequality, concurrence, quantum discord for a two-qubit
bipartite system is an exhaustive way of analyzing the inseparability along with quantumness of the system,
but there is an applicational aspect to it.  One of such application is teleportation.

In the early nineties  a new aspect of quantum entanglement was discovered, viz.
teleportation \cite{ben}. Teleportation is purely based on classical information and
non-classical Einstein-Podolsky-Rosen (EPR) correlations \cite{epr}. The basic scheme 
of teleportation
is to transfer an arbitrary quantum state from sender to receiver using a pair of particles in a singlet state shared by them. When two parties 
share a mixed state  instead of a singlet,  it is not possible to teleport an unknown quantum state with full fidelity. The natural question 
was  to find out the optimal value of  teleportation fidelity of an unknown quantum state;
a fidelity above which will ensure  non-classical character of the state forming
the quantum channel. It was shown that for a purely classical channel the optimum
teleportation fidelity is $F=\frac{2}{3}$ \cite{gis,hor,mas}.

In Ref. \cite{pop},  the following  question regarding quantum teleportation,
Bell-CHSH (Clauser-Horne-Shimony-Holt) inequalities  and inseparability was raised: What is the exact relation between Bell
inequalities violation and teleportation?
Bell inequalities are basically built upon the locality and reality assumption and have
nothing to do with quantum mechanics. It is quite obvious that the state which violates Bell's inequality can be useful for teleportation. However,  in another
work, Werner \cite{wer} gave an example of an entangled state, which has the unique feature of
not violating Bell's inequality for a certain range of classical probability of mixing.  An interesting question to ask  was whether 
states which do not violate Bell-CHSH inequalities
are suitable for teleportation.  An answer to this question was offered  in the 
form of examples of mixed two spin-$\frac{1}{2}$ entangled states that do not violate
Bell-CHSH inequalities but  can still be useful for teleportation \cite{hor,chak,adh}. However any state that violates Bell-CHSH inequalities is always suitable for teleportation.  In a recent work \cite{adh1},  
these relations for  two qubit mixed entangled state arising from the Buzek Hillery quantum cloning machine were probed.

An issue of central importance is the study of quantum correlations in mixed states.  Open quantum systems  \cite{bp02} provide a 
natural setting for a systematic discussion of mixed states.  A
basic motivation of this work is to find out whether there are mixed, entangled states (having  positive concurrence) which do not violate Bell's inequality but 
can still  be useful for teleportation.  We consider  entangled states  generated by open system models involving
quantum non demolition (purely dephasing) as well as  dissipative  type of system-reservoir interaction. 
In addition, we make a comparative study of various measures of quantum correlations like concurrence, 
Bell's inequality, quantum discord and teleportation fidelity  on mixed states, generated by the above evolutions. 
Interest  in  the  relevance  of open  system  ideas  to  quantum
information has  increased in recent  times because of  the impressive
progress made on the experimental front in the manipulation of quantum
states of  matter towards  quantum information processing  and quantum
communication.  Myatt {\it et  al.} \cite{myatt} and Turchette {\it et
  al.} \cite{turch}  have performed a  series of experiments  in which
they  simulated both pure dephasing as well as dissipative evolutions,  
by coupling  the  atom  (their
system-$S$) to  engineered reservoirs, in  which the coupling  to, and
the  state  of,  the  environment  are  controllable.  

There has been considerable interest, in recent times, in the study of quantum correlations in
open quantum systems.   Quantum and classical correlations have been
studied, in the context of quantum phase transitions \cite{ms09}. An operational measure of quantum correlations, based on the
dynamics of these correlations in  open system evolutions, was proposed in \cite{mv09}, while
in \cite{ms10}, a study of classical and quantum correlations was made, on a two-qubit
system interacting with two independent environments.  A study of quantum discord and entanglement
was made, on a system consisting of two coupled quantum dots in \cite{fc1},  and it was shown
that discord can be more resistant to dissipation that entanglement. In 
\cite{wb09},  Markovian dynamics,  and in \cite{fc2},  non-Markovian
dynamics of quantum discord was made on a two qubit system,  and it was seen that, in some 
conditions,  quantum discord and entanglement can behave very differently. Interesting experimental
investigations have also been made in this context.  In \cite{ps10}, a theoretical and experimental study
was made on the dynamics of classical and quantum correlations in an NMR quadrupolar system, while  in
\cite{xg10},  the dynamics of different kinds of bipartite correlations was investigated, experimentally, in an
all-optical setup.  In \cite{kp10}, an interesting experiment was presented, in which dissipation induces entanglement 
between two atomic objects, thereby paving the way for long-lived entanglement, in a steady state. In a recent work, it was shown that almost all quantum states have a non zero quantum discord \cite{fer}. In reference \cite{qas}, a numerical comparison between discord and entanglement of formation was made. Some authors investigate the dynamical relations among entanglement, mixedness and non locality, quantified by concurrence C, purity P and maximum of Bell function B, respectively, in a system of two qubits in a common structured reservoir \cite{mazz}. In another work, the dynamics of quantum and classical correlations in presence of non dissipative decoherence is studied \cite{mazz1}.

In another important work \cite{cav} the authors a general characterization of quantum discord and entanglement in large
families of states (Bell Diagonal states) has been presented. This also gives a kind of a general analysis of the structure of entanglement and discord in particular of their non analytic behaviour under decoherence. However, in this work we also provide a general analysis of the various measures of quantum correlations for the states that are modeled by
a   two-qubit system interacting with its environment
via a quantum non demolition (purely dephasing) as well as dissipative type of  interaction. 

The basic motivation of the paper is  three fold:

 1) The primal objective is to  classify the  set of two qubit states modeled by a two qubit system interacting with its environment 
via a quantum non demolition as well as dissipative type of interaction. We categorize  these density matrices into three broad divisions \cite{mw10}:

 i) \textit{Entangled states}: The density matrices having positive concurrence.

 ii) \textit{Non classical separable states}: The states with a non zero discord and zero concurrence (absence of entanglement).

 iii) \textit{Classical states}: The  separable states with zero discord.

 In this process of classification of the generated states our aim is to give a systematic overview of the threshold values of the 
dynamical parameters at which the transition from one class of states to other class is taking place.

 2) Next, our motivation to study  teleportation fidelity and Bell's inequality of the states, generated by this open system model, 
is to find out  entangled states which do not violate Bell's inequality but can still  be useful for teleportation. 

 3) Last but not the least our objective is to see the interface of  quantum and classical correlations in the presence of  
dissipative decoherence. In other words, it remains interesting to study the classical as well as quantum dechorence  in presence as well as absence of entanglement.

The plan of the paper  is as follows.  We discuss  various measures, both from a fundamental as well as an applicational perspective,
of quantum correlations and try to bring out their interconnections.  
Two-qubit  systems, used here in our study of quantum correlations in mixed states,  are  then discussed, briefly,
from the perspective of a general division of open quantum systems into a purely dephasing (non demolition) or dissipative evolution. This is then
applied to a study of quantum correlations,  followed by our discussions and conclusions.

\section{An Overview of Measures of Quantum Correlation: Bell's Inequality, Concurrence, Discord and Teleportation Fidelity}
In this section we try to analyze different measures of quantum correlation and make a comparative study between them.  Among the measures 
studied are Bell's inequality,  concurrence  and quantum discord, from a theoretical perspective. 
In addition, we also discuss teleportation fidelity as a measure,
from an applicational point of view.

\subsection{Bell Inequalities}

Bell's inequality was one of the first tools used to detect entanglement. Originally, Bell inequalities
were introduced as an attempt to rule out local hidden variable (LHV) models.

Consider a bipartite system of two qubits where Alice and Bob share a particle, each supplied and initially prepared by another party, say, Charlie. 
Each of them are allowed to perform measurements on their respective particle. Once Alice receives her particle she performs a measurement on it. 
Alice is provided with two sets of measurement operators and she could choose to do one of the two measurements. 
These are labeled by $P_{M1}$ and $P_{M2}$, respectively. 
Since Alice does not know in advance which measurement to apply, she adopts a random method to make her decision. 
Let us assume that each of these measurements can have two possible values $\{+1,-1\}$.  Let $M1$ and $M2$ be the values revealed by the two measurements  $P_{M1}$ and $P_{M2}$. Similarly,  Bobs measurements are labeled by $P_{M3}$ and $P_{M4}$.  Each of these ${M1}$, ${M2}$, ${M3}$ and ${M4}$ can have the values $\{+1,-1\}$. 
Bob does not decide in advance which measurement he will carry out and waits until he has received the particle from Alice and then chooses randomly. The setup is so arranged that they carry out their measurements in a causally connected manner. Thus the principle of no signalling ensures that the measurement of one particle cannot affect the measurement of the other.

Let us now consider the  algebraic expression $(M1)(M3)+(M2)(M3)+(M2)(M4)-(M1)(M4)$. 
Since $M1,M2=\pm 1$, it follows in either case that $(M1)(M3)+(M2)(M3)+(M2)(M4)-(M1)(M4)=\pm 2$.    
Now if consider the probability distribution and calculate the mean value of the above expression, then a little calculation gives  the 
standard form of the Clauser-Horne-Shimony-Holt inequality \cite{chs}
\begin{eqnarray}
E[(M1)(M3)+(M2)(M3)+(M2)(M4)\nonumber\\-(M1)(M4)]=E[(M1)(M3]+E[(M2)(M3]+\nonumber\\E[(M2)(M4]-E[(M1)(M4)]\leq 2,
\end{eqnarray} 
where $E$ stands for the mean value.
Interestingly,  it can be seen that in standard quantum theory, it is always possible to design experiments for which this inequality gets violated \cite{asp1,asp2,asp3}. 
This shows that quantum physics violates local realism, which implies that these results cannot be described by a LHV model. 
It may also provoke the implication that, if  measurements on a quantum state violate a Bell's inequality,  the state is entangled. 
However, with the advent of the Peres-Horodecki criteria \cite{per,hor1},  the converse of this statement need not be true.

One can express the most general form of Bell-CHSH inequality for the mixed state $\rho=\frac{1}{4}[I\otimes I+(r.\sigma)\otimes I+I\otimes (s.\sigma)
\nonumber\\+\sum_{n,m=1}^3 t_{mn}(\sigma_m\otimes \sigma_n)] $ as $M(\rho)<1$, where $M(\rho)=max(u_i+u_j)$,  $u_i,u_j$ 
are the eigenvalues of the matrix $T^{\dagger}T$ (where the elements of the correlation matrix $T$ is given by, $t_{mn}=Tr[\rho(\sigma_m\otimes \sigma_n)]$ and $T^{\dagger}$ is the conjugate transpose of $T$) \cite{hor}.  
As mentioned above,  violation of Bell's inequality for a given quantum state indicates that the state is entangled. 
But at the same time, there are certain entangled states which do not violate Bell's inequality.  

\subsection{Concurrence}

Since Bell's inequality is not  able to detect all possible entangled states,  there is a need for some kind of measure which will quantify the amount of entanglement 
present in the system.  A well known measure of entanglement is
concurrence which for a two-qubit system, is equivalent
to the entanglement of formation. The concurrence
of a pure two-qubit state $|\psi_i\rangle$ is given by
\begin{eqnarray}
 C(|\psi_i\rangle)=\sqrt{2(1-Tr\rho_A^2)}=\sqrt{2(1-Tr\rho_B^2)},
\end{eqnarray}
where $\rho_A = Tr_B|\psi\rangle\langle \psi|$ is the partial trace of $|\psi\rangle\langle \psi|$ over
subsystem B, and $\rho_B$ is the subsystem obtained when we trace out A. For a mixed
state $\rho$,  concurrence is  defined as the optimization of average concurrence
of   all pure state  decompositions of 
 $\rho=\sum_j p_j|\psi_j\rangle\langle \psi_j|$,
\begin{eqnarray}
 C(\rho)=min\sum_jp_jC(|\psi_j\rangle).
\end{eqnarray}

For a mixed state $\rho$ of two qubits, concurrence is \cite{woo}
\begin{eqnarray}
 C=max(\lambda_1-\lambda_2-\lambda_3-\lambda_4,0),
\end{eqnarray}
where $\lambda_i$ are the square root of the eigenvalues,  in decreasing order, 
of the matrix $\rho^{\frac{1}{2}}(\sigma_y\otimes \sigma_y)\rho^*(\sigma_y\otimes \sigma_y)\rho^{\frac{1}{2}}$.
$\rho^*$ denotes  complex conjugation of $\rho$ in the computational basis
$\{|00\rangle, |01\rangle, |10\rangle, |11\rangle\}$ and $\sigma_y$ is the Pauli operator. 
The entanglement of formation (EOF) can then be expressed as a monotonic function of concurrence C as
\begin{eqnarray}
 E_F=-\frac{1+\sqrt{1-C^2}}{2}\log_2(\frac{1+\sqrt{1-C^2}}{2})\nonumber\\-\frac{1-\sqrt{1-C^2}}{2}\log_2(\frac{1-\sqrt{1-C^2}}{2}).
\end{eqnarray}
Thus we see that concurrence, as a function of EOF,  is a  measure of  quantum correlation as it actually quantifies the amount of entanglement present in the system.

\subsection{Quantum Discord}

As noted above,  entanglement and quantum correlations  need not be  identical. 
Quantum discord attempts to measure and quantify all quantum correlations including entanglement \cite{oll,hen,luo}.

In classical information theory, the correlation between two random variables $X$ and $Y$ is given by a quantity called `Mutual Information' $J(X:Y)=H(X)-H(X|Y)$, where $H(X|Y)$ is the conditional entropy of $X$ given that $Y$ has already occurred and $H(X)$ is the Shannon entropy of the random variable $X$.  Since $H(X|Y)=H(X,Y)-H(Y)$, 
there exists an alternative 
expression for mutual information  $I(X:Y)=H(X)+H(Y)-H(X,Y)$. Classically,  there is no ambiguity between these two expressions of mutual information and they are same.  But a real difference arises in the quantum regime. In the quantum case, let us consider a bipartite state $\rho_{XY}$, where $\rho_X$ and $\rho_Y$ 
are the states of the individual subsystems. Here we note that in the quantum case, Shannon entropies
$H(X), H(Y)$ are replaced by von-Neumann entropies (e.g: $H(X)=H(\rho_X)=-Tr_X\rho_XLog(\rho_X)$). Now, from the definition itself the conditional entropy $H(X|Y)$
requires a specification of the state of $X$ given the state of $Y$.  Such a statement in quantum theory is ambiguous until the 
to-be-measured set of states of  $Y$ are selected. For that reason we focus on perfect measurements of Y defined by a set of one dimensional
projectors $\{\pi^Y_j\}$. The subscript $j$ is used for indexing 
different outcomes of this measurement.
The state of X, after the measurement is given by
\begin{eqnarray}
\rho_{X|\pi^Y_j}=\frac{\pi^Y_j \rho_{XY} \pi^Y_j}{Tr(\pi^Y_j \rho_{XY})},
\end{eqnarray}
with probability $p_j={Tr(\pi^Y_j \rho_{XY})}$. Thus, $H(\rho_{X|\pi^Y_j})$ is the von-Neumann entropy of the system in the state $\rho_X$, given that  projective measurement 
is carried out on  system $Y$ in the most general basis $\{\cos(\theta)|0\rangle+\exp(i\phi)\sin(\theta)|1\rangle,\exp(-i\phi)\sin(\theta)|0\rangle-\cos(\theta)|1\rangle\}$. 
The entropies $H(\rho_{X|\pi^Y_j})$ weighted by the probabilities $p_j$, yield  the conditional entropy
of X,  given the complete set of measurements $\{\pi^Y_j\}$
 on Y,  as $H(X|\{\pi^Y_j\})=\sum_j p_jH(\rho_{X|\pi^Y_j})$.  From this, the quantum analogue of $J(X:Y)$ is seen to be 
\begin{eqnarray}
J(X:Y)=H(X)-H(X|\{\pi^Y_j\}),
\end{eqnarray}
while $I(X:Y)$ is similar to its classical counterpart
\begin{eqnarray}
I(X:Y)=H(X)+H(Y)-H(X,Y).
\end{eqnarray}
It is clearly evident that these two expressions are not identical in  quantum theory. Quantum discord is
the difference between these two generalizations of classical mutual information,
\begin{eqnarray}
D(X:Y)=H(Y)-H(X,Y)+H(X|\{\pi^Y_j\}).
\end{eqnarray}
 The classical correlation is given by the difference between the mutual information and quantum correlation
\begin{eqnarray}
C(X:Y)=I(X:Y)-D(X:Y). 
\end{eqnarray}

We thus see  from the above  expression, that quantum discord aims to quantify the amount of quantum correlation 
that remains in the system and also points out that classicality and separability are not synonymous. In other words,  it actually reveals the quantum advantage over the classical correlation. 

\subsection{Teleportation Fidelity}
In addition to all these  measures of quantum correlation one could also
attempt to quantify them in terms of an application, for e.g.,  fidelity of teleportation \cite{ben}. 
It involves the separation of an input state into its classical and
quantum part from which the state can be reconstructed
with perfect fidelity $F = 1$. The basic idea is to use a pair
of particles in a singlet state shared by sender (Alice) and
receiver (Bob).  Popescu \cite{pop} noticed that  pairs in a mixed state could be still useful for (imperfect) teleportation. 
Consider the following  general representation of the mixed state of a two-qubit system :
\begin{eqnarray}
\rho &=& \frac{1}{4}[I\otimes I+(r.\sigma)\otimes I\nonumber\\ &+& I\otimes (s.\sigma)+\sum_{n,m=1}^3 t_{mn}(\sigma_m\otimes \sigma_n)] ,
\end{eqnarray}
where $\rho$ acts on Hilbert space $\mathcal{H}=\mathcal{ H}_1\otimes\mathcal{ H}_2$,
and $I$ stands for identity operator.  Here $\mathcal{H}_1$, $\mathcal{H}_2$ are the individual Hilbert spaces of the
two qubits and $\sigma_{n}$  with $n = 1, 2, 3$ are the standard
Pauli matrices, r, s are vectors in $R^3$,  $r.\sigma = \sum_{i=1}^3 r_i\sigma_i$.
The quantities $t_{nm} = Tr[\rho (\sigma_n \otimes\sigma_m)]$ are the coefficients of a real matrix denoted  by T. 
This representation is most convenient when one talks about the inseparability of  mixed states.  In fact,  
all  the parameters fall into two different classes:  those that  describe the local behavior of the state, i.e., 
($r$ and $s$),  and those responsible for correlations (T matrix). 

In the standard teleportation scheme we have a mixed state $\rho$ acting as a quantum channel (originally formed by pure singlet states \cite{ben}). 
One of the particles is with Bob while the other one and a third
particle in an unknown state $|\phi\rangle$ are subjected to 
joint measurement in Alice's Hilbert space. These measurement operators are given by a family of projectors
\begin{eqnarray}
 P_k=|\psi_k\rangle\langle \psi_k|,k=0,1,2,3,
\end{eqnarray} 
where $\psi_k$ constitute the so-called Bell basis. Then using two
bits Alice sends  Bob the result of outcome k on basis of which he 
applies some unitary transformation $U_k$, obtaining
in this way his particle in a state k.
Then the fidelity of  transmission of the unknown
state is given by \cite{hor},
\begin{eqnarray}
 F=\int_S dR(\phi)\sum_k p_kTr(\rho_kP_{\phi}),,
\end{eqnarray}
where the integral is taken over all states (indexed by the angle $\phi$) belonging to
the Bloch sphere with uniform distribution $R$ and $p_k =
Tr [(P_k\otimes I)(P_{\phi} \otimes \rho )]$ denotes the probability of the k-th
outcome. Now the task is to find those unitary transformations $U_k$ that produce
the highest fidelity (a choice of a quadruple of such $U_k$ is what would be called a  strategy). 
Maximizing F over all strategies gives \cite{hor}
\begin{eqnarray}
F_{max}&=&\frac{1}{2}(1+\frac{1}{3}N(\rho))\nonumber\\&=&\frac{1}{2}(1+\frac{1}{3}[\sqrt{u_1}+\sqrt{u_2}+\sqrt{u_3}]).
\end{eqnarray}
Here $u_i$ and $u_j$ are
the eigenvalues of $U=T^{\dag}(\rho)T(\rho)$,
where $T(\rho)=[T_{ij}],T_{ij}=Tr[\rho(\sigma_i\otimes \sigma_j)]$ and $T^{\dag}$ implies the conjugate transpose of $T$. 
The classical fidelity of teleportation in the absence of entanglement is  obtained as $\frac{2}{3}$. 
Thus whenever $F_{max}>\frac{2}{3} (N(\rho)>1)$,  teleportation is possible.

At this point it is quite interesting to note that there is a non-trivial interplay between Bell's inequality and  
teleportation fidelity. This is because both $M(\rho), N(\rho)$ are dependent on the correlation matrix T. The 
relationship between these two quantities can be well understood in the form of the inequality $N(\rho)>M(\rho)$.  
Hence, it is clear that  states which do violate Bell's inequality are always useful for teleportation. However,  interestingly,  
this  does not rule out the possibility of  existence of entangled states that do not violate Bell's inequality, but can 
still be useful for teleportation. Indeed, we discuss below examples of such states.

\section{A Brief Introduction to Open Quantum Systems: Quantum  Non Demolition and Dissipative Type}
Open  quantum systems  is  the study  of  evolution of  the system  of
interest  taking into  account its  interaction with  the environment (reservoir or bath).
Let the  total
Hamiltonian be $H = H_S +  H_R + H_{SR}$  , where $S$ stands  for the
system,  $R$  for the  reservoir  and  $SR$  for the  system-reservoir
interaction.  The evolution  of the system of interest  $S$ is studied
taking into  account the  effect of its  environment $R$,  through the
$SR$    interaction    term,    making    the    resulting    dynamics
non-unitary.
Based upon the system-reservoir  interaction, open quantum systems can
be  broadly   classified  into  two  categories:   (A).   Quantum  Non
Demolition (QND),  where $[H_S, H_{SR}] = 0$  resulting in decoherence
without dissipation and (B).  Dissipative, with $[H_S, H_{SR}] \neq 0$
resulting in  decoherence along with dissipation \cite{bg07}.   Our model, for
the study of quantum correlations, 
is a two-qubit system interacting with its environment, 
envisaged as a bath of non interacting harmonic oscillators,  via QND or
dissipative  type of  interactions.  
This provides the quantum channel used here, as in for e.g.,  teleportation,  to study quantum correlations.
The  qubit-bath  interactions are
position  dependent  which leads  to  interesting collective  behavior
that can  be broadly classified into independent (localized)  or collective dynamical
regimes.  Below we discuss the model briefly, details of which can be
found in \cite{bs1, bs2}.

\subsection{Two-Qubit QND Interaction with a Squeezed Thermal Bath}
We  consider the Hamiltonian,  describing the  QND interaction  of two
qubits with the bath as \cite{bg07, rj02, bgg07}
\begin{eqnarray}
H & = & H_S + H_R + H_{SR} \nonumber \\ & = & \sum\limits_{n=1}^2 \hbar 
\varepsilon_n S^z_n+ \sum\limits_k \hbar \omega_k b^{\dagger}_k
b_k \nonumber\\ & & + \sum\limits_{n,k} \hbar S^z_n  (g^n_k b^{\dagger}_k + 
g^{n*}_k b_k). \label{basich} 
\end{eqnarray} 
Here  $H_S$, $H_R$  and $H_{SR}$  stand  for the  Hamiltonians of  the
system,  reservoir  and  system-reservoir  interaction,  respectively.
$b^{\dagger}_k$, $b_k$ denote  the creation and annihilation operators
for the  reservoir oscillator of frequency  $\omega_k$, 
$S^z_n$ is the energy operator of the $n$th qubit,
$g^n_k$ stands
for the coupling  constant (assumed to be position  dependent) for the
interaction  of the  oscillator  field with  the  qubit system and is
taken to be
\begin{equation}
g^n_k = g_k e^{-ik.r_n}, \label{coupling}
\end{equation}
where  $r_n$ is  the  qubit position.   Since  $[H_S, H_{SR}]=0$,  the
Hamiltonian (1) is of QND type. In the parlance of quantum information
theory,  the  noise  generated  is  called  the  phase  damping  noise
\cite{gp06, bg07}.

The  position dependence of  the coupling  of the  qubits to  the bath
(\ref{coupling}) helps to bring out the effect of entanglement between
qubits through the qubit separation: $ r_{mn} \equiv r_m - r_n$.  This
allows  for  a  discussion  of  the  dynamics  in  two  regimes:  (A).
independent decoherence where $k.r_{mn} \sim \frac{r_{mn}}{\lambda} \geq
1$   and   (B).    collective   decoherence   where   $k.r_{mn}   \sim
\frac{r_{mn}}{\lambda}  \rightarrow  0$. The  case  (B) of  collective
decoherence would arise  when the qubits are close  enough for them to
experience  the  same  environment,  or  when  the  bath  has  a  long
correlation  length  (set   by  the  effective  wavelength  $\lambda$)
compared to  the interqubit separation $r_{mn}$  \cite{rj02}.  Our aim
is to study the reduced dynamics  of the qubit system. 
We assume separable initial conditions, i.e., 
\begin{equation}
\rho (0) = \rho^s (0) \otimes \rho_R (0), \label{initial}
\end{equation}
where
\begin{equation}
\rho^s (0) = \rho^s_1(0) \otimes \rho^s_2(0), \label{sysinitial}
\end{equation}
is the initial state of the qubit system and the subscripts
denote the  individual qubits.~ In  Eq.  (\ref{initial}),
$\rho_R (0)$ is  the initial density matrix of  the reservoir which we
take  to be  a   broadband   squeezed thermal
bath \cite{bg07,bgg07,gp06} given by
\begin{equation}
\rho_R(0) = S(r, \Phi) \rho_{th} S^{\dagger} (r, \Phi), \label{rhorin}
\end{equation}
where
\begin{equation}
\rho_{th} = \prod_k \left[ 1 - e^{- \beta \hbar \omega_k} 
\right] e^{-\beta \hbar \omega_k b^{\dagger}_k b_k} \label{rhoth}
\end{equation}
is the  density matrix  of the thermal  bath at temperature  $T$, with
$\beta \equiv 1/(k_B T)$, $k_B$ being the Boltzmann constant, and
\begin{equation}
S(r_k, \Phi_k) = \exp \left[ r_k \left( {b^2_k \over 2} e^{-2i 
\Phi_k} - {b^{\dagger 2}_k \over 2} e^{2i \Phi_k} \right) 
\right] \label{sqop}
\end{equation}
is  the  squeezing  operator   with  $r_k$,  $\Phi_k$  being  the
squeezing parameters \cite{cs85}.  In order to obtain the reduced  dynamics of the system, 
we trace over the reservoir  variables \cite{bs1}. The results pertaining to a thermal reservoir
without squeezing can be obtained by setting the squeezing parameters to zero.

\subsection{Two-Qubit Dissipative Interaction with a Squeezed Thermal Bath}
We consider the Hamiltonian, describing the dissipative interaction of
$N$ qubits (two-level atomic system)  with the bath (modeled as a 3-D
electromagnetic field (EMF)) via the dipole interaction as \cite{ft02}
\begin{eqnarray}
H & = & H_S + H_R + H_{SR} \nonumber \\ & = & \sum\limits_{n=1}^N \hbar 
\omega_n S^z_n + \sum\limits_{\vec{k}s} \hbar \omega_k (b^{\dagger}_{\vec{k}s}
b_{\vec{k}s} + {1 / 2}) \nonumber\\ &-& i\hbar \sum\limits_{\vec{k}s}
\sum\limits_{n=1}^N [\vec{\mu}_n . \vec{g}_{\vec{k}s} (\vec{r}_n)(S_n^+ + 
S_n^-)b_{\vec{k}s}- h.c.]. \label{hamiltonian} 
\end{eqnarray}
Here $\vec{\mu}_n$ are the transition dipole moments, dependent on the
different atomic positions $\vec{r}_n$ and 
\begin{equation}
S_n^+ = | e_n \rangle \langle g_n|,~ S_n^- = | g_n \rangle \langle e_n|,
\label{dipole}
\end{equation}
are  the dipole raising  and lowering  operators satisfying  the usual
commutation relations and
\begin{equation}
S_n^z = \frac{1}{2}(| e_n \rangle \langle e_n| -  |g_n \rangle \langle g_n|),
\label{energy}
\end{equation}
is    the    energy    operator    of   the    $n$th    atom,    while
$b^{\dagger}_{\vec{k}s}$,   $b_{\vec{k}s}$   are   the  creation   and
annihilation operators  of the field  mode (bath) $\vec{k}s$  with the
wave vector $\vec{k}$, frequency  $\omega_k$ and polarization index $s
= 1,2$ with the system-reservoir (S-R) coupling constant
\begin{equation}
\vec{g}_{\vec{k}s} (\vec{r}_n) = (\frac{\omega_k}{2 \varepsilon \hbar V})^
{1/2} \vec{e}_{\vec{k}s} e^{i \vec{k}.r_n}. \label{coupling1}
\end{equation}
Here  $V$  is   the  normalization  volume and $\vec{e}_{\vec{k}s}$ is 
the unit polarization vector of the field.   It  can   be  seen  from
Eq. (\ref{coupling1})  that the S-R  coupling constant is  dependent on
the  atomic position  $r_n$. This  leads  to a  number of  interesting
dynamical aspects, as seen below.  From now we will concentrate on the
case of two qubits.  Assuming separable initial conditions,  and taking
a trace over  the bath,  in a squeezed thermal state,  
the reduced density matrix  of the qubit system
in the interaction picture and in the usual Born-Markov, rotating wave
approximation (RWA) is  \cite{ft02}
\begin{eqnarray}
\frac{d\rho}{dt} &=& -\frac{i}{\hbar}[H_{\tilde{S}}, \rho] - 
\frac{1}{2} \sum\limits_{i,j=1}^2 \Gamma_{ij}[1+\tilde{N}] \nonumber\\ 
& & \times (\rho S_i^+ S_j^- +  S_i^+ S_j^- \rho 
- 2 S_j^- \rho  S_i^+)\nonumber\\
&-& \frac{1}{2} \sum\limits_{i,j=1}^2 \Gamma_{ij}\tilde{N} 
(\rho S_i^- S_j^+ +  S_i^- S_j^+ \rho - 2 S_j^+ \rho  S_i^-) \nonumber\\
&+& \frac{1}{2} \sum\limits_{i,j=1}^2 \Gamma_{ij}\tilde{M} 
(\rho S_i^+ S_j^+ +  S_i^+ S_j^+ \rho - 2 S_j^+ \rho  S_i^+)\nonumber\\
&+& \frac{1}{2} \sum\limits_{i,j=1}^2 \Gamma_{ij}\tilde{M^*} 
(\rho S_i^- S_j^- +  S_i^- S_j^- \rho  \nonumber\\ 
&-& 2 S_j^- \rho  S_i^-). \label{red}
\end{eqnarray}
In Eq. (\ref{red}) 
\begin{equation}
\tilde{N} = N_{\rm th}(\cosh^2(r) + \sinh^2(r)) + \sinh^2(r), \label{N} 
\end{equation}
\begin{equation}
\tilde{M} = -\frac{1}{2} \sinh(2r) e^{i\Phi} (2 N_{\rm th} + 1)
\equiv Re^{i\Phi(\omega_0)}, \label{M}
\end{equation}
with 
\begin{equation}
\omega_0 = \frac{\omega_1 + \omega_2}{2},    \label{omega0}
\end{equation}
and
\begin{equation}
N_{\rm th} = {1 \over e^{{\hbar \omega \over k_B T}} - 1}. \label{sqpara}
\end{equation}
Here  $N_{\rm th}$  is the  Planck distribution  giving the  number of
thermal  photons  at  the  frequency  $\omega$  and  $r$,  $\Phi$  are
squeezing parameters.   The analogous case  of a thermal  bath without
squeezing can be obtained from  the above expressions by setting these
squeezing parameters  to zero, while setting the  temperature ($T$) to
zero one recovers the case of the vacuum bath.   Here 
the assumption of perfect matching of the squeezed
modes  to the  modes of  the  EMF is  made along  with, the  squeezing
bandwidth  being much larger  than the  atomic line widths.   Also, the
squeezing carrier frequency is taken to be tuned in resonance with the
atomic frequencies.

In Eq. (\ref{red}),
\begin{equation}
H_{\tilde{S}} = \hbar \sum\limits_{n=1}^2 \omega_n S^z_n + \hbar
\sum^2_{\stackrel{i,j}{(i \neq  j)}} \Omega_{ij} S_i^+ S_j^-, \label{unitary}
\end{equation}
where 
\begin{eqnarray}
\Omega_{ij}   &=&   \frac{3}{4}   \sqrt{\Gamma_i   \Gamma_j}\left[-[1-
(\hat{\mu}.\hat{r}_{ij})^2]\frac{\cos(k_0  r_{ij})}{k_0  r_{ij}}  \right.   \nonumber\\  &+&   \left.  [1-
3(\hat{\mu}.\hat{r}_{ij})^2]  
[\frac{\sin(k_0    r_{ij})}{(k_0     r_{ij})^2}    +    \frac{\cos(k_0
r_{ij})}{(k_0 r_{ij})^3}]\right].               \label{omegacol}      
\end{eqnarray}
Here $\hat{\mu}  = \hat{\mu}_1  = \hat{\mu}_2$ and  $\hat{r}_{ij}$ are
unit  vectors   along  the   atomic  transition  dipole   moments  and
$\vec{r}_{ij}  = \vec{r}_i  - \vec{r}_j$,  respectively.  Also  $k_0 =
{\omega_0  / c}$,  with $\omega_0$  being as  in  Eq.  (\ref{omega0}),
$r_{ij} = |\vec{r}_{ij}|$.  The wave vector $k_0 = {2\pi / \lambda_0}$,
$\lambda_0$ being  the resonant wavelength, occurring in  the term $k_0
r_{ij}$ sets  up a  length scale into  the problem depending  upon the
ratio  ${r_{ij} /  \lambda_0}$. This  is  thus the  ratio between  the
inter atomic  distance  and the  resonant  wavelength,  allowing for  a
discussion  of   the  dynamics  in  two   regimes:  (A).   independent
decoherence  where $k_0.r_{ij}  \sim \frac{r_{ij}}{\lambda_0}  \geq 1$
and    (B).    collective    decoherence   where    $k_0.r_{ij}   \sim
\frac{r_{ij}}{\lambda_0}  \rightarrow 0$. The  case (B)  of collective
decoherence would arise  when the qubits are close  enough for them to
feel the  bath collectively  or when the  bath has a  long correlation
length (set  by the resonant wavelength $\lambda_0$)  in comparison to
the interqubit separation $r_{ij}$.  $\Omega_{ij}$ (\ref{omegacol}) is
a collective coherent effect due to the multi-qubit interaction and is
mediated via the bath through the terms
\begin{equation}
\Gamma_i   =  \frac{\omega^3_i   \mu^2_i}{3   \pi  \varepsilon   \hbar
c^3}. \label{single}
\end{equation}
The  term $\Gamma_i$  is  present  even in  the  case of  single-qubit
dissipative system  bath interaction \cite{bsri06, bsdiss}  and is the
spontaneous emission rate, while
\begin{equation}
\Gamma_{ij} = \Gamma_{ji} = \sqrt{\Gamma_i \Gamma_j} F(k_0 r_{ij}),
\label{gammacol}
\end{equation}
where $i \neq j$ with
\begin{eqnarray}
F(k_0 r_{ij})   &=&   \frac{3}{2} \left[[1-
(\hat{\mu}.\hat{r}_{ij})^2]\frac{\sin(k_0  r_{ij})}{k_0  r_{ij}} +  [1-
3(\hat{\mu}.\hat{r}_{ij})^2]   \right.   \nonumber\\  &\times&   \left.
[\frac{\cos(k_0    r_{ij})}{(k_0     r_{ij})^2}    -    \frac{\sin(k_0
r_{ij})}{(k_0 r_{ij})^3}]\right].               \label{fcol}      
\end{eqnarray}
$\Gamma_{ij}$ (\ref{gammacol}) is the collective incoherent effect due
to the dissipative multi-qubit interaction with the bath. For the case
of identical qubits, as  considered here, $\Omega_{12} = \Omega_{21}$,
$\Gamma_{12} = \Gamma_{21}$ and $\Gamma_1 = \Gamma_2 = \Gamma$.

\section{Quantum correlations in Open Quantum Systems}

Here we study quantum correlations, in two-qubit mixed states, generated by QND as well as
dissipative system-reservoir interactions,  as discussed above.  We give examples 
of entangled states that do not violate Bell's inequality, but  can be useful for teleportation with a fidelity better than the classical fidelity 
of teleportation.  On these states, we made a comparative study of various features of quantum correlation like teleportation fidelity $(F_{max})$, 
violation of Bell's inequality $M(\rho)$ (violation takes place for $M(\rho) \geq 1$),  
concurrence $C(\rho)$ and discord with respect to various experimental parameters like, 
bath squeezing parameter $r$,  inter-qubit spacing $r_{12}$, temperature $T$ and time of evolution $t$. 
Since these  states are very close to Werner like states and  since these Werner like states can be written in any  basis, 
we measure the discord only in the computational basis $\{|0\rangle,|1\rangle\}$.  Werner states being basis independent,  
the value of  discord  remains same in any basis.  The motivation of this work is basically three fold. First of all, our motivation is to  classify the  set of two qubit states generated by a realistic open system models. We classify these density matrices into three broad categories: 1) Entangled states having positive concurrence,  2) Non classical separable states with a non zero discord and zero concurrence (entanglement) and 3) classical states i.e separable states with zero discord. In this process of classification of the generated states we also give a systematic overview of the threshold values of the dynamical parameters at which the transition of one category of states to the other category  is taking place.  Also, we  look  for  realistic open system models that  generate  entangled states which can be useful for 
teleportation, but at the same time,  not violate Bell's inequality. Finally in presence of dissipative decoherence we make a comparative study of the quantum and classical correlation. Interestingly we find that after a certain time the rate of quantum as well as classical dechoerence becomes identical.  An analogous study for non dissipative systems was made in \cite{mazz1}.

\subsection{Quantum Correlations in a QND System}

Here we study examples of two-qubit density matrices obtained as a result of QND evolution, in both the collective as well as  
independent (localized) regimes of the model. Teleportation fidelity $(F_{max})$, 
violation of Bell's inequality $M(\rho)$, concurrence $C(\rho)$ and discord are examined for these states as a function of  various  parameters like: 
bath squeezing parameter $r$, inter-qubit spacing $r_{12}$, temperature $T$ and  system-bath interaction time $t$. 

 We categorize these density matrices into three broad categories: 1) entangled states, 2) separable non classical states and 3) classical states. We specifically mention the range of the dynamical parameters for which such a  classification is possible.  
We  also report  cases where these two qubit states are entangled states, but do not violate Bell's inequality and still can be useful for teleportation with a fidelity better than the classical fidelity of teleportation.


 The first example considers   two-qubit density matrices, from the  collective model,   as illustrated in Figs.  (\ref{fig1}a),  (b),  (c) and (d),  
where the behavior of correlation measures:
concurrence,  teleportation fidelity,  Bell's inequality violation and discord, respectively,  are studied 
with respect to the bath squeezing parameter $r$.  As seen from Fig. (\ref{fig1}(c)), $M(\rho)<1$ for all values of the parameter, indicating that in the range shown,
the states satisfy Bell's inequality and hence can  be described by a local realistic model. Interestingly, for this range of $r$, 
Fig. (\ref{fig1}(b)) shows that  teleportation fidelity is greater than $\frac{2}{3}$, clearly indicating that the state is useful for teleportation. 
This is corroborated by
Fig. (\ref{fig1}(a)),  where a plot of concurrence shows that concurrence is positive (indicating  presence of entanglement), 
attaining a maximum value $0.13$,  approximately.
Figure (\ref{fig1}(d)),  shows  non-vanishing discord for the above range of the parameter $r$. Since both concurrence and discord attains  non zero values for all range of the bath squeezing parameter $r$, it indicates that the two qubit density matrices obtained in this collective model is entangled for all values of $r$ in this range.   It can also be seen that, 
for the value of the various parameters chosen, 
with increase in bath squeezing, i.e., with increase in the
absolute value of the parameter $r$,  concurrence, teleportation fidelity, $M(\rho)$ decrease, but discord can increase.

 The next example,  depicted in  Figs.  (\ref{fig2}(a)), (b), (c), (d),   studies two-qubit density matrices,  from the independent  model. 
In  Fig. (\ref{fig2}(a)), concurrence is plotted with respect to bath squeezing parameter $r$.  It is seen  that  states are entangled when $r$ lies in the range $[-1.66,1.66]$.
Teleportation fidelity, as in the Fig. (\ref{fig2}(b)),  indicates that the states are useful for teleportation for the same range of $r$, i.e.,  when they are entangled.
However, from Bell's inequality, as shown  in  Fig. (\ref{fig2}(c),  
in the same range, we see that the states do not violate Bell's inequality.
Interestingly, from Fig. (\ref{fig2}(d)),  we find a non zero quantum discord in the range $[-3,3]$ and particularly,  in the range $[-3,-1.66]\cup[1.66,3]$, i.e.,  where
entanglement as depicted in  Fig. (\ref{fig2}(a)) is zero,  discord is non-zero. Thus we conclude that the states obtained in this independent model are non classical separable states  for the above range of the bath squeezing parameter $r$.  This brings out  the fact that  the amount of entanglement present 
in a system is not equivalent to the  the total amount of quantum correlation in it.  For the parameters chosen, we see a similar pattern of the various correlation functions
as in the previous figure, i.e., with the exception of discord, the quantum correlation measures fall with the increase in bath squeezing.

 Our last example from a QND evolution,  in the independent regime of the model,  is as shown in
Figs.  (\ref{fig3}(a)), (b), (c), (d), where  concurrence,  teleportation fidelity,  Bell's inequality and discord, respectively, are depicted with respect to the
inter-qubit spacing $r_{12}$.
In  Fig.  (\ref{fig3}(a)), from the  plot of concurrence,  it can be seen that the states are entangled with a positive concurrence for all values of $r_{12}$, 
except in the range $(1.4,1.7),(4.6,4.8)$. Like in the previous example, we see  non-vanishing  value of discord in the range  where there is no entanglement, as shown  in  Fig. (\ref{fig3}(d)). The states generated in this range are not entangled but they are non classical separable states.

 Figure (\ref{fig3}(b)) shows  that the states  useful for teleportation are from the same range of $r_{12}$,  for which they are entangled.
From  Fig. (\ref{fig3}(c)),  it can be seen that there are certain regions where Bell's inequality is violated $r_{12} \in [2.9,3.1]$, but mostly it is satisfied.
.

\subsection{Quantum Correlations in Dissipative Systems}

Here we consider those two-qubit density matrices which are generated as a result of dissipative system-bath interaction.  A 
comparative analysis of various features of quantum correlations, such as,  concurrence $C(\rho)$, teleportation fidelity $(F_{max})$, 
violation of Bell's inequality $M(\rho)$ and discord,  is made on these states, generated by the above density matrices, 
as a function of  various  parameters: bath squeezing parameter $r$, inter-qubit spacing $r_{12}$, temperature $T$ and time of system-bath evolution $t$. 

 We classify the density matrices generated as a result of dissipative system-bath interaction into three categories:  1) entangled 
states, 2) separable non classical states and 3) classical states. Quite similar to the analysis of the  non dissipative systems here also we provide the threshold values of the dynamical parameters for which the transition from one particular class to another particular class is taking place. 
We also find examples of states which are entangled but do not violate Bell's inequality and still can be useful for teleportation with a fidelity better 
than the classical fidelity of teleportation.
Finally we study the overall dynamics of quantum and classical correlations in presence of dissipative decoherence.

 For our first example, we study the behavior of correlation measures:
concurrence,  teleportation fidelity,  Bell's inequality violation and discord, as illustrated in Figs. (\ref{fig4}a),  (b),  (c) and (d), respectively, 
as a function of the inter-qubit distance $r_{12}$. 
From  Fig.  (\ref{fig4}(a)),  we find that the two qubit density matrix is entangled with a positive concurrence for $r_{12} < 0.5$.
Moreover,  from  Fig.  (\ref{fig4}(d)),   a positive discord is seen for the complete range of $r_{12}$, even in the range where there is no entanglement ($r_{12} \geq 0.5$). Interestingly, we see that discord attains a constant value $.2544$ in this range. We conclude from this non zero value of  discord even in the absence of entanglement that the states in this range are non classical separable states.   
Figure (\ref{fig4}(b)) illustrates  that $F_{max}>\frac{2}{3}$, for all values of $r_{12}$ except  where there is no entanglement. In the absence of entanglement the teleportation fidelity attains the constant value $\frac{2}{3}$.
However, from  Fig. (\ref{fig4}(c)) we find that $M(\rho)<1$ for all values of $r_{12}$ and attains a constant value $0.478$ in the absence of entanglement. This  clearly demonstrates that these states can be useful for teleportation despite the fact that they satisfy Bell's inequality.

As a function of the  inter-qubit distance,  the various correlation measures exhibit oscillatory behavior, in the collective regime of the model, 
but flatten out subsequently
to attain almost constant values in the independent regime of the model.  This oscillatory behavior is due to the strong collective behavior 
exhibited by the dynamics due to the relatively close proximity of the qubits in the collective regime \cite{bs1,bs2}.

Next, we study the behavior of quantum correlations,  in  two-qubit density matrices in the  collective regime of the model, with respect to the evolution time $t$.
In  Fig.  (\ref{fig5}(a)),  concurrence is seen to exhibit  damped oscillations. Here we see that the states are entangled initially until the time evolution parameter $t$ reaches the value $0.371$. 
Figure  (\ref{fig5}(b)),  for  teleportation fidelity $F_{max}$, also shows a  damped  behavior and can be seen, in general,  to be greater than $\frac{2}{3}$, 
except for the range $t \geq 0.371$ where there is no entanglement (zero concurrence), where it attains a constant value  $\frac{2}{3}$.
From Fig. (\ref{fig5}(c)),  we find that the states satisfy Bell's inequality for all values $t$. 
Discord, as in Fig. (\ref{fig5}(d)), is always positive, though its value is decreasing with time. We conclude from the figure that the states that are generated  in the collective regime are initially entangled but after a certain value of the time evolution parameter $t=0.371$, the states produced are no longer entangled but  are non classical separable states. The threshold value of the discord at $t=0.371$ is $0.183$.

 In Fig. (\ref{fig6}), we have an example of mixed states, belonging to the collective regime of the dissipative two-qubit model,
where concurrence is zero for all values of the parameter $r$.  These states satisfy
Bell's inequality, as is evident from  (\ref{fig6}(c)) (it takes a constant value $.338$), but teleportation fidelity, Fig. (\ref{fig6}(b)), is equal to  $2/3$, which is
consistent with our understanding that states that are not entangled cannot be used for teleportation. However, 
these states have a discord (\ref{fig6}(d))  of about $0.127$. So basically these states  are non classical separable states.

In Figs. (\ref{fig7} (a), (b)) we have examples of mixed states ,  belonging to the collective and independent regimes of the two-qubit model, respectively. We study explicitly the evolution of the quantum correlation (quantum discord), quantum mutual information, concurrence and classical correlation with the passage of time. Here we see that unlike non dissipative dynamics \cite{mazz1},  no transition from classical to quantum dechorence takes place. On the contrary both classical and quantum correlations are lost due to the interaction with the environment. Interestingly we find that in the absence of entanglement from a certain time $t>\bar{t}$, the classical correlation and the quantum discord becomes identical. This cleary indicates that after a certain time  the rate of quantum and classical decoherence becomes same.

\section{Discussion}
 Here  we provide the result of our study in a nutshell.
Our first objective was to classify the states. In the following table we give the classification of the states in the context of various dynamical parameters.

~~~~~~~~~~~~~~~~~~~~~~~~~~~~~~~~~~~~~\underline{{\bf TABLE I:}}

\begin{tabular}{|c|c|c|c|}
\hline  Figures  & Entangled States & Non Classical  & Classical States \\    &  & Separable States & \\
\hline Fig 1 & Yes  & No & No \\ & $\forall r$ & &\\
\hline Fig 2 & Yes  & Yes & Yes\\
 & $r \in[-1.66,1.66]$ & $r \in [-3,-1.66]$ & $r \in [-5,-3]$\\&&$\cup[1.66,3]$&$\cup[3,5]$\\
\hline Fig 3 & Yes & Yes & No\\
 & $r_{12} \in[1,1.4]$ & $r_{12} \in (1.4,1.7)$ & \\&$\cup[1.7,4.6]$&$\cup(4.6,4.8)$&\\&$\cup[4.8,5]$&& \\
\hline Fig 4 & Yes & Yes & No\\
 & $r_{12}\in [0.1, 0.5)$ & $r_{12} \in [0.5,2]$ & \\
\hline Fig 5 & Yes & Yes & No\\
 & $t\in [0.1, 0.371)$ & $t \in [0.371,1.1]$ & \\
\hline Fig 6 & No & Yes & No\\
 &  & $\forall r$ & \\
\hline
\end{tabular}\\

 Our next motivation was to provide examples of entangled states that do not violate Bell's inequality, but  can still be useful for 
teleportation. We summarize our findings in the following table. 

~~~~~~~~~~~~~~~~~~~~~~~~~~~~~~~~~~~~~\underline{{\bf TABLE II:}}

\begin{tabular}{|c|c|c|}
\hline  Figures  & Bells Inequality Violation  & $(F_{max}>\frac{2}{3})$ \\&$M(\rho)>1$&\\    
\hline Fig 1 & No  & Yes \\
\hline Fig 2 & No  & Yes $r \in[-1.66,1.66]$\\&& No $r \in [-5,-1.66]$\\&& $\cup[1.66,5]$\\
 \hline Fig 3 & Yes $r_{12} \in [2.9,3.1]$ & Yes $r_{12} \in[1,1.4]$\\&&$\cup[1.7,4.6]$\\&&$\cup[4.8,5]$\\& No $r_{12} \in
[1,2.9)$ & No $r_{12} \in(1.4,1.7)$ \\&$\cup(3.1,5]$&$\cup (4.6,4.8)$\\
 \hline Fig 4 & No & Yes $r_{12}\in [0.1, 0.5)$\\&& No $r_{12}\in [0.5, 2]$\\
\hline Fig 5 & No & Yes $t\in [0.1, 0.371)$\\&& $t\in [0.371, 1.1]$\\
\hline Fig 6 & No & No\\
\hline
\end{tabular}\\

 Finally we studied the dynamics of  quantum as well as classical correlation for states generated by dissipative open system 
interactions. From Figs. (\ref{fig7} (a), (b)) we obtain certain features that are unlike the QND model, that is,  there is no sudden transition between quantum and classical loss of correlations in a composite system. On the contrary the rate of  loss of correlations are same in both the cases after certain time.\\
\section{Conclusion}
 In this work, we have tried to understand the nature of quantum correlations in mixed states.
We focus on two-qubit systems, and  the mixed states we work with are generated by concrete 
Open System models, with  system-bath interactions
such that both  pure dephasing (QND) as well as dissipative evolutions are obtained.

 Our primal objective was to classify the states, obtained as a result of both  pure dephasing (QND) as well as dissipative 
evolutions into three categories: 1) Entangled states,  2) Non Classical Separable states, 
3) Classical states. We were able to make this classification in the context of various dynamical parameters for both QND and dissipative systems.
 
 One of our objectives was to see whether one can generate states, from realistic models, which satisfy Bell's inequality, that is, 
can be described by a local realistic theory, yet can be used for
a useful task employing entanglement, viz. teleportation.   We were able to give a number of such examples, both for the QND as well
as  dissipative type of evolutions.

 In addition our objective was to see the dynamics of the quantum as well as classical correlations in the presence of dissipative 
decoherence. Analyzing the dynamics, we conclude that there do exist certain class of states obtained as result of a dissipative evolution for which the rate of decay of classical and quantum correlations are identical after a certain time interval. 

These models, as well as the general evolutions considered, can be envisaged in an experimental setup \cite{myatt, turch, ft02}.
This puts our study of quantum correlations on a firm basis.

\begin{widetext}
\begin{center}
\begin{figure}
\qquad
\subfloat[]{\includegraphics[height=4.5cm,width=7.159cm]{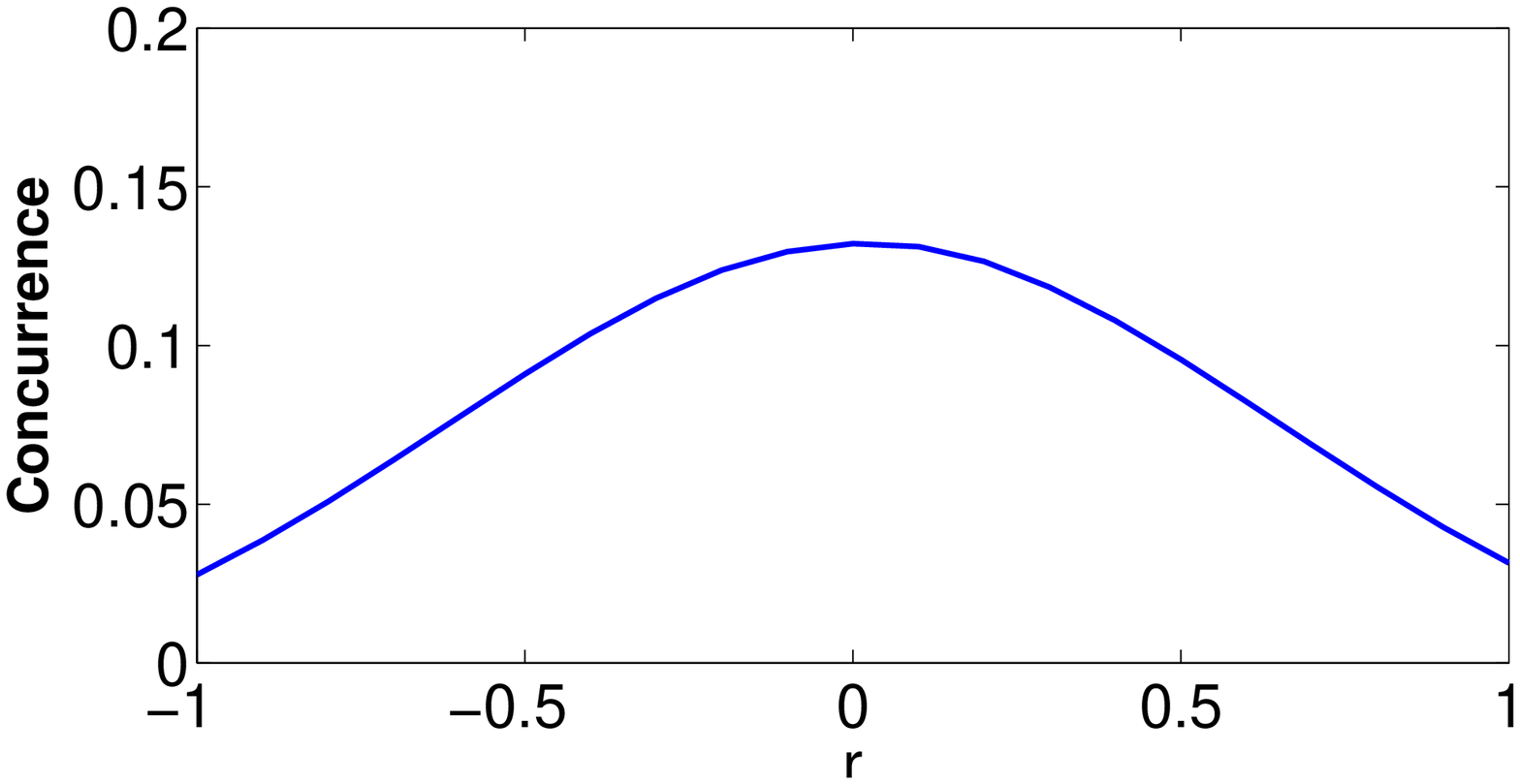}}
\subfloat[]{\includegraphics[height=4.5cm,width=8cm]{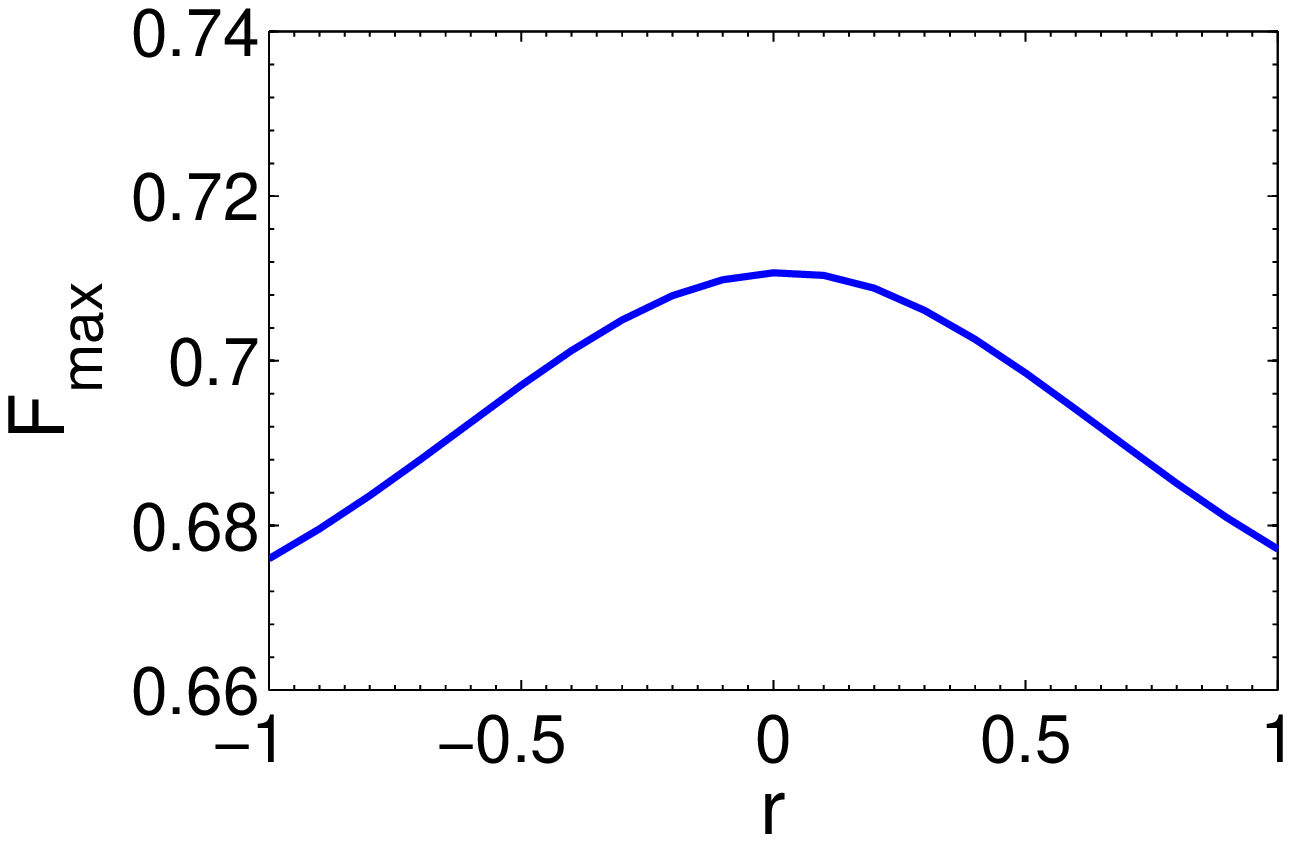}}\\
\subfloat[]{\includegraphics[height=4.5cm,width=8cm]{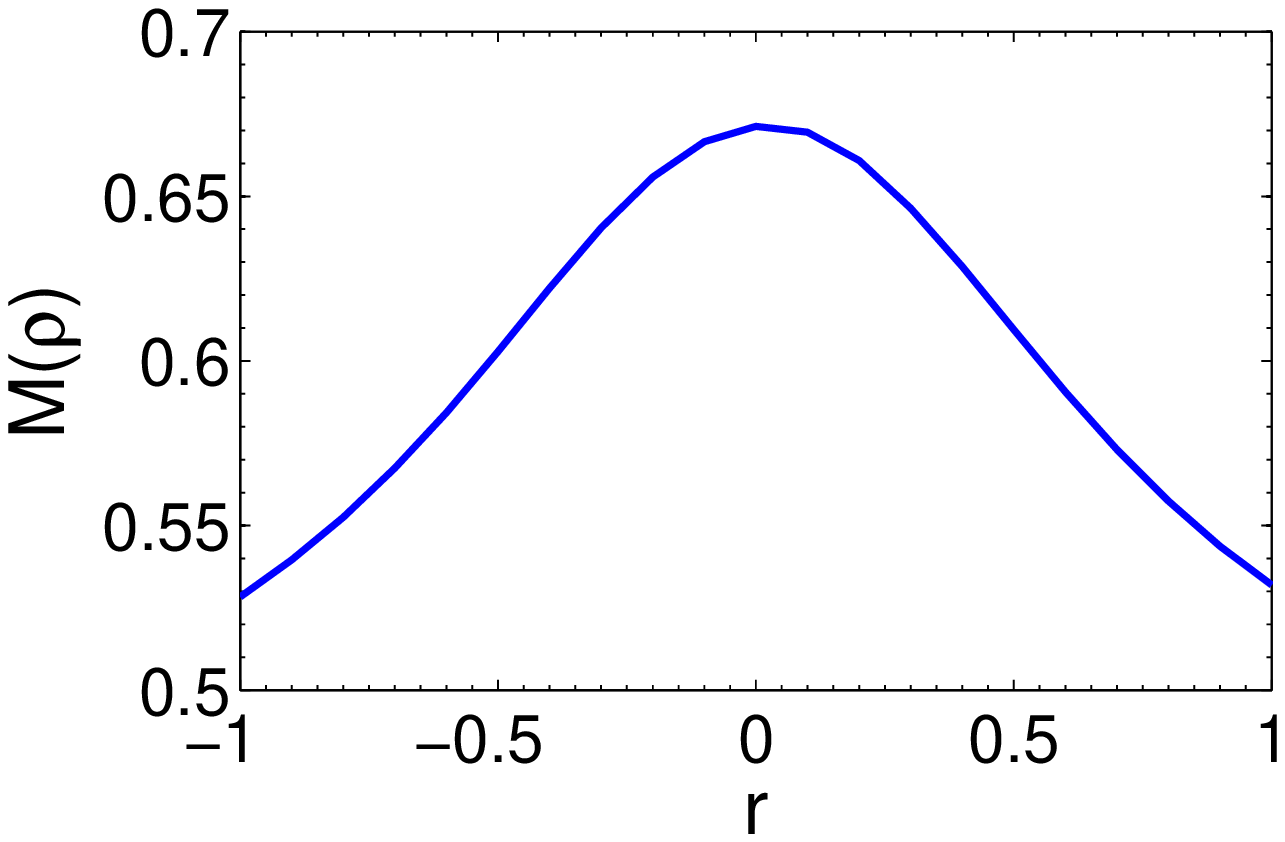}}
\quad \subfloat[]{\includegraphics[height=4.5cm,width=7.19cm]{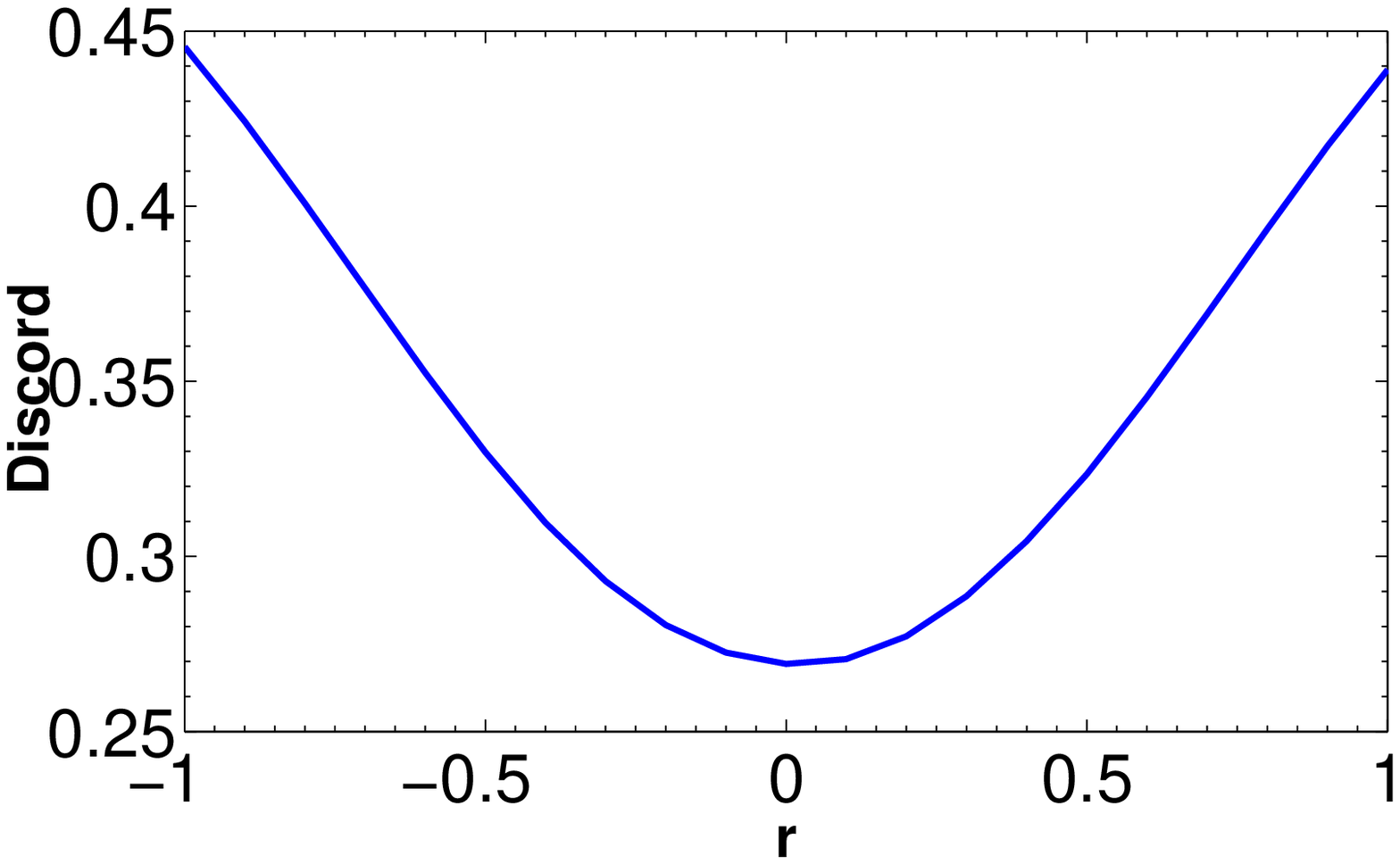}}\\
\caption{Quantum correlations in a two-qubit system undergoing a QND
interaction.  The Figs. (a), (b), (c) and (d) represent the evolution of
concurrence, maximum teleportation fidelity $F_{max}$, test of Bell's inequality $M(\rho)$,
discord as a function of bath squeezing parameter $r$.  Here 
temperature  $T$ (in  units where $\hbar\equiv  k_B=1$)  is 25, evolution time $t$ is 1 and the 
inter-qubit distance $r_{12} = 0.05$, i.e.,  the  collective  decoherence  model.}\label{fig1}
\end{figure}

\begin{figure}
\subfloat[]{\includegraphics[height=4.0cm,width=7.0cm]{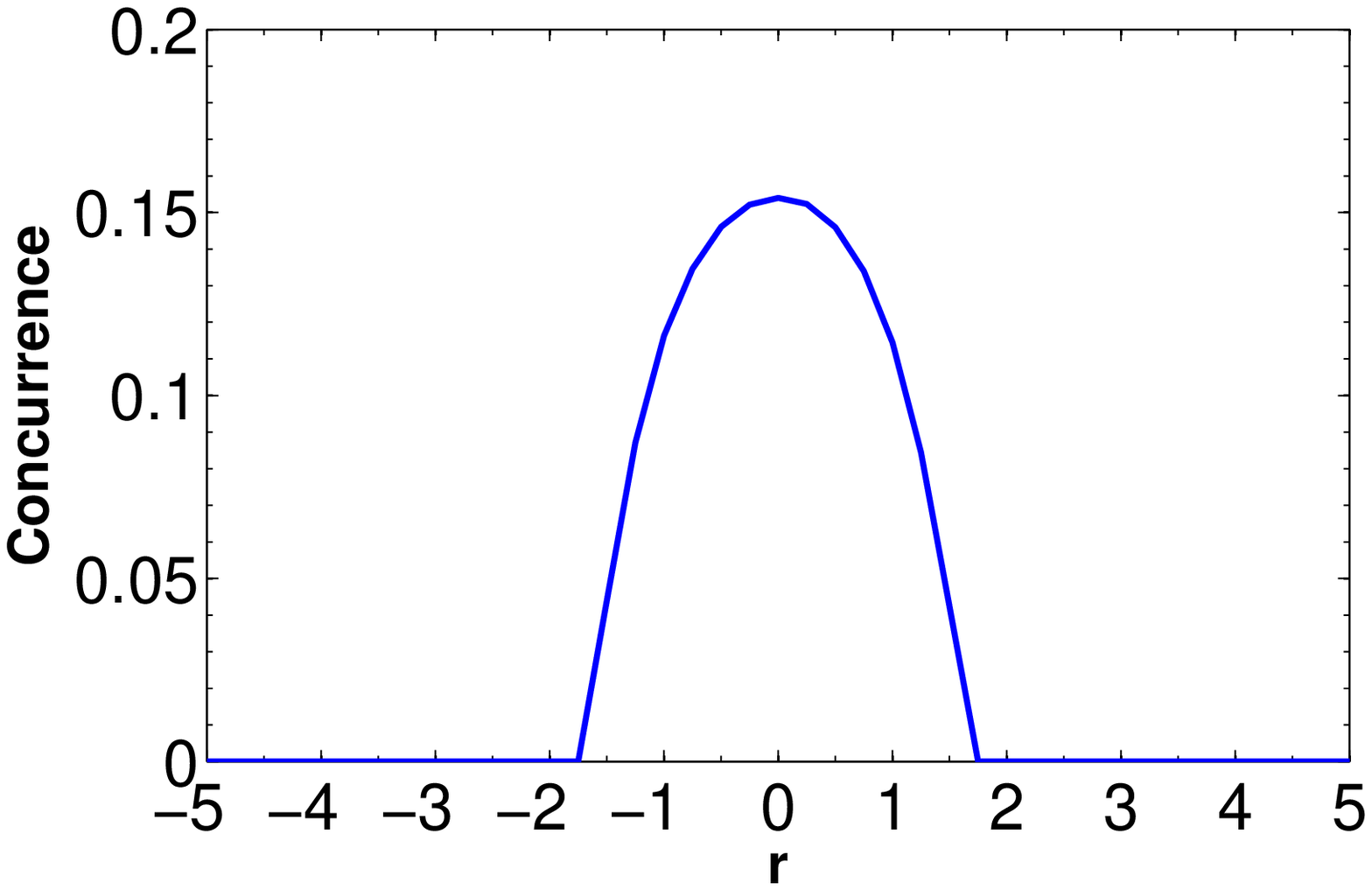}}
\subfloat[]{\includegraphics[height=4.0cm,width=7.5cm]{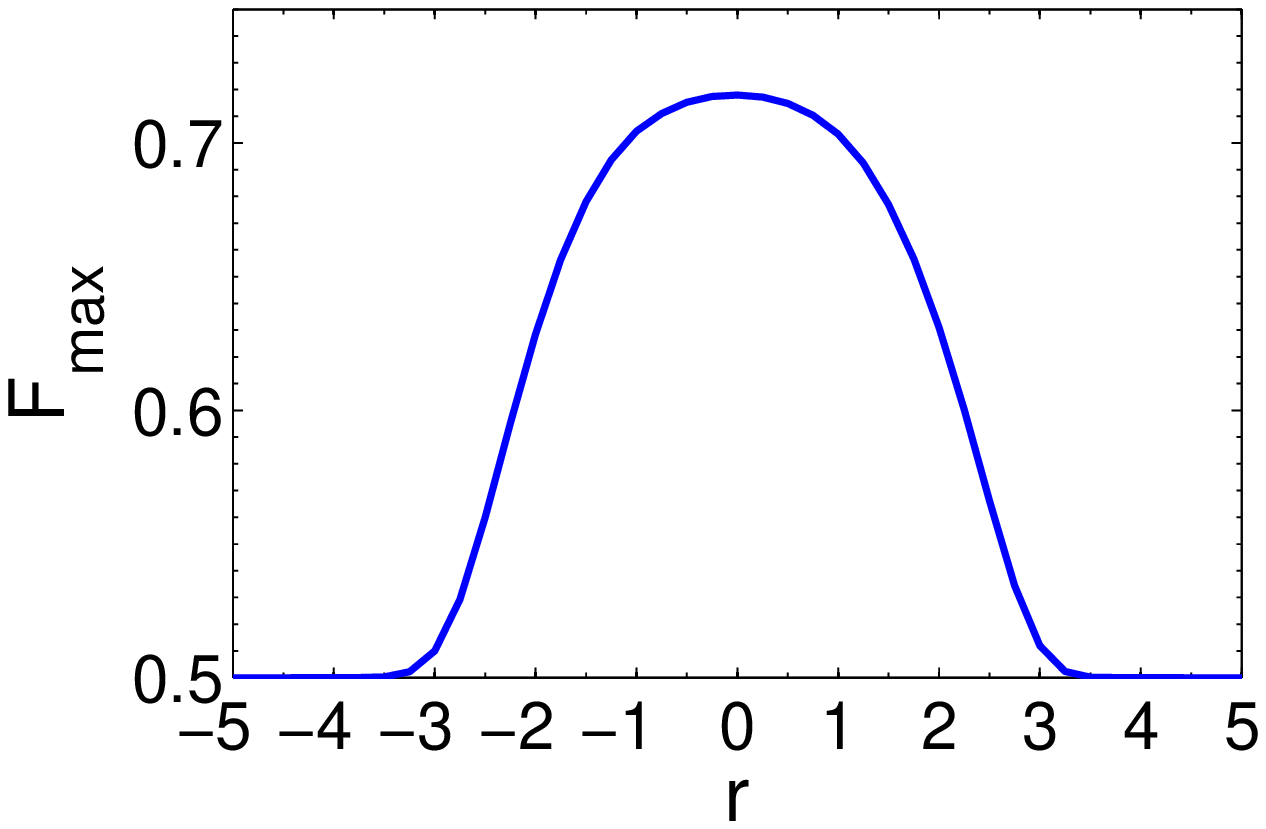}}\\
\subfloat[]{\includegraphics[height=4.0cm,width=7.1cm]{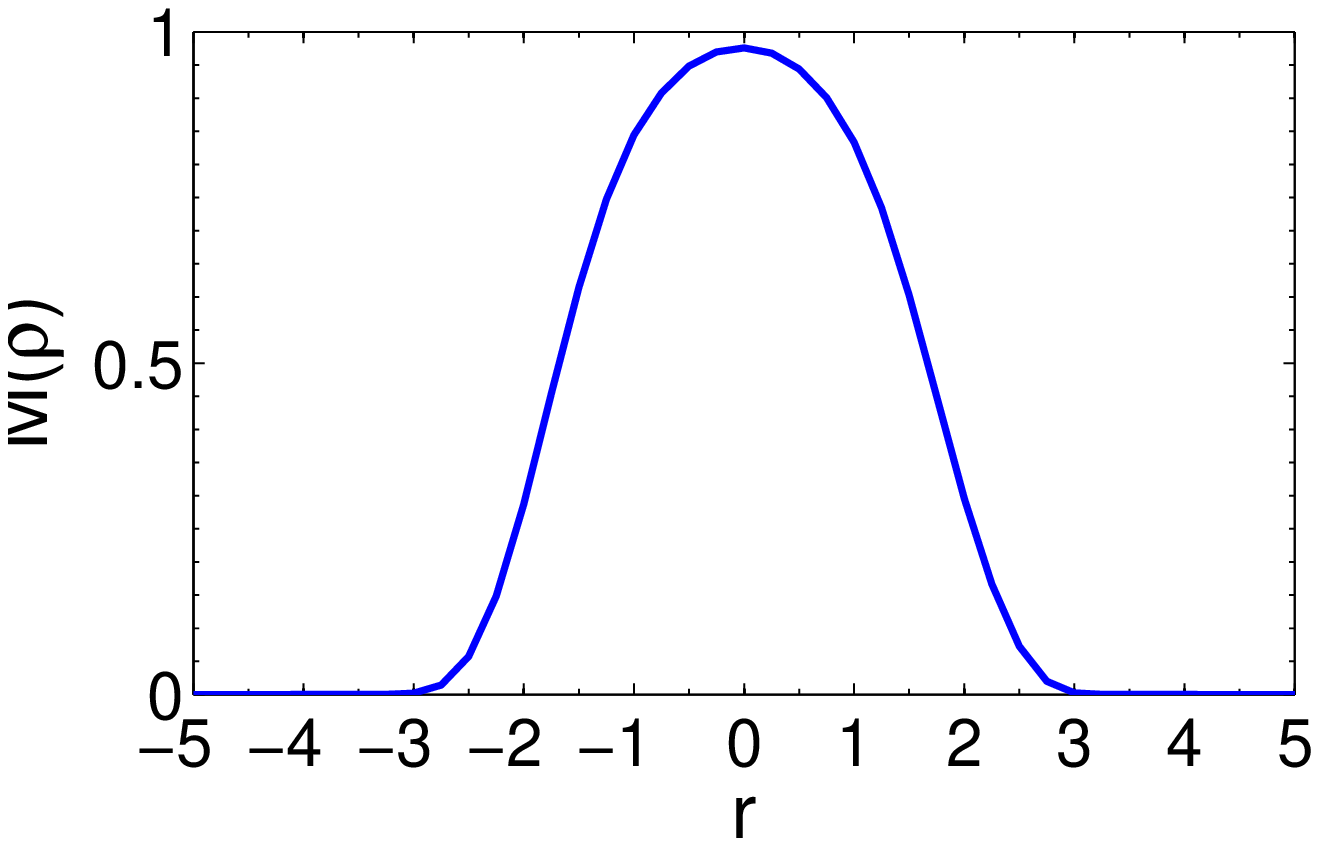}}
\quad \subfloat[]{\includegraphics[height=3.8cm,width=7.0cm]{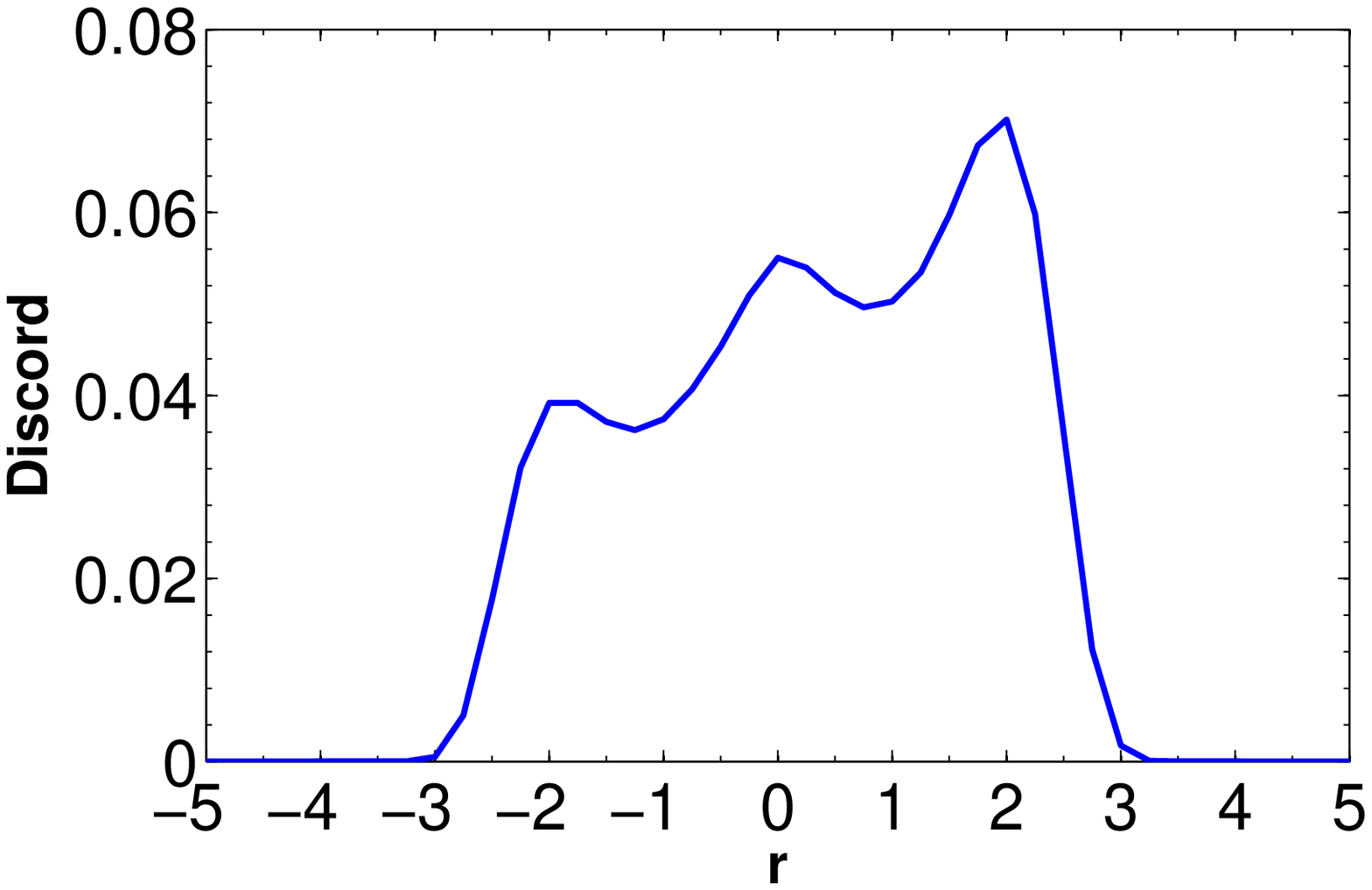}}\\
\caption{Figures (a), (b), (c) and (d) represent the evolution of
concurrence, maximum teleportation fidelity $F_{max}$, test of Bell's inequality $M(\rho)$, and
discord, respectively,  as a function of bath squeezing parameter $r$,  in a two-qubit system evolving via a QND
interaction. Here 
temperature  $T = 0$, evolution time $t$ is 1.1 and the 
inter-qubit distance $r_{12} = 1.05$, i.e.,  the  independent decoherence  model.}\label{fig2}
\end{figure}

\begin{figure}
\subfloat[]{\includegraphics[height=4.0cm,width=7.0cm]{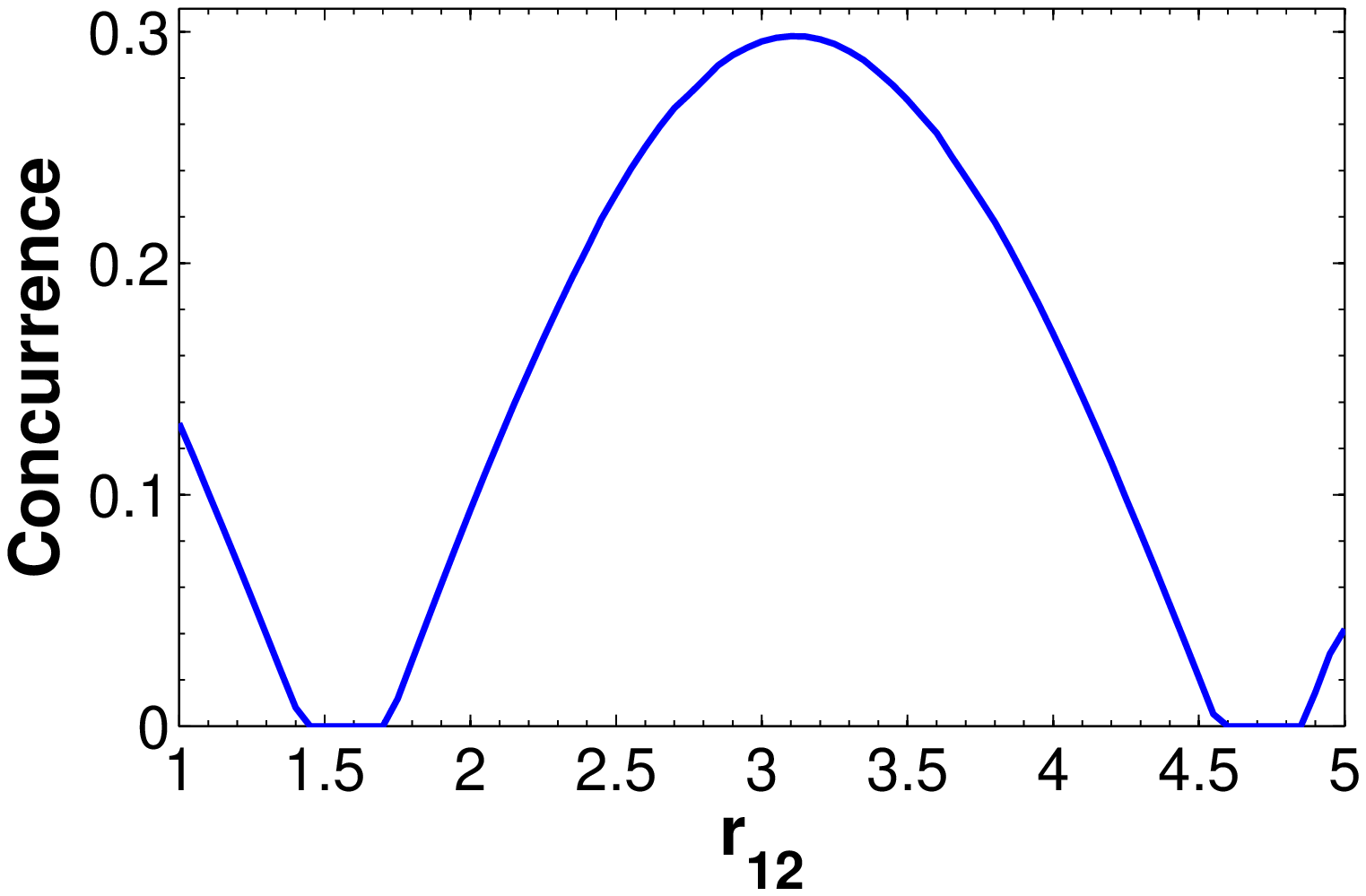}}
\subfloat[]{\includegraphics[height=4.0cm,width=7.0cm]{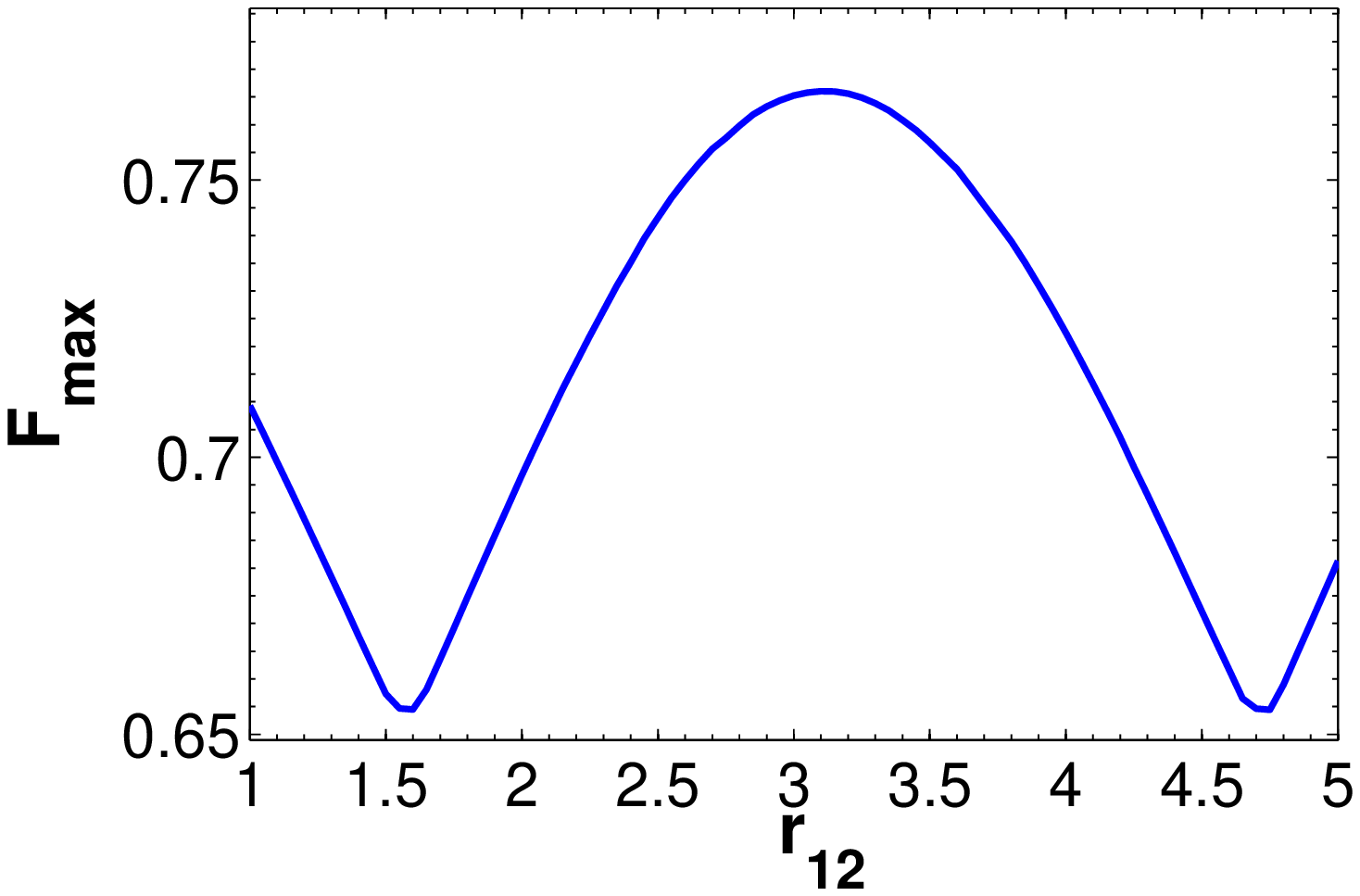}}\\
\subfloat[]{\includegraphics[height=4.0cm,width=7.0cm]{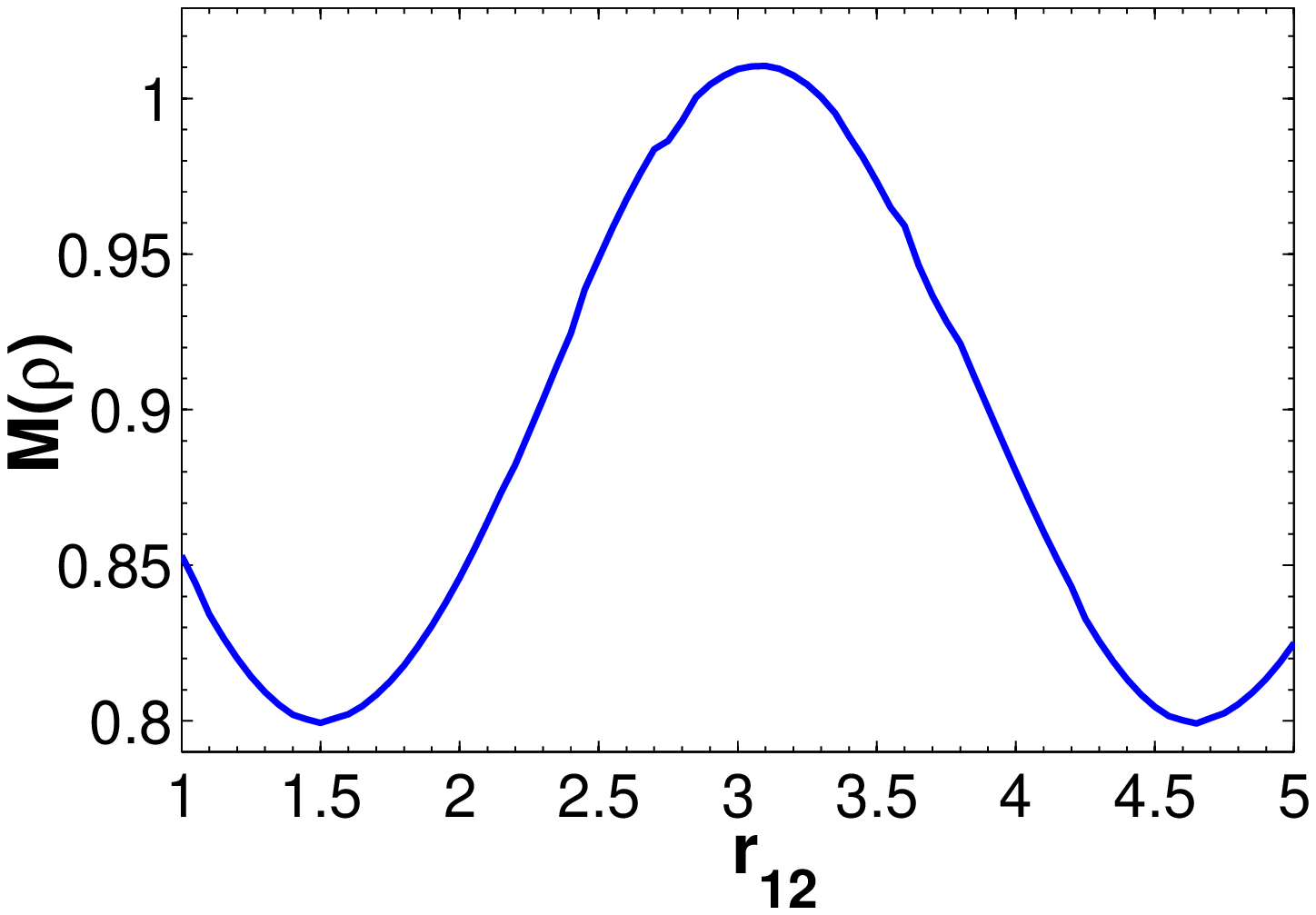}}
\subfloat[]{\includegraphics[height=4.0cm,width=7.0cm]{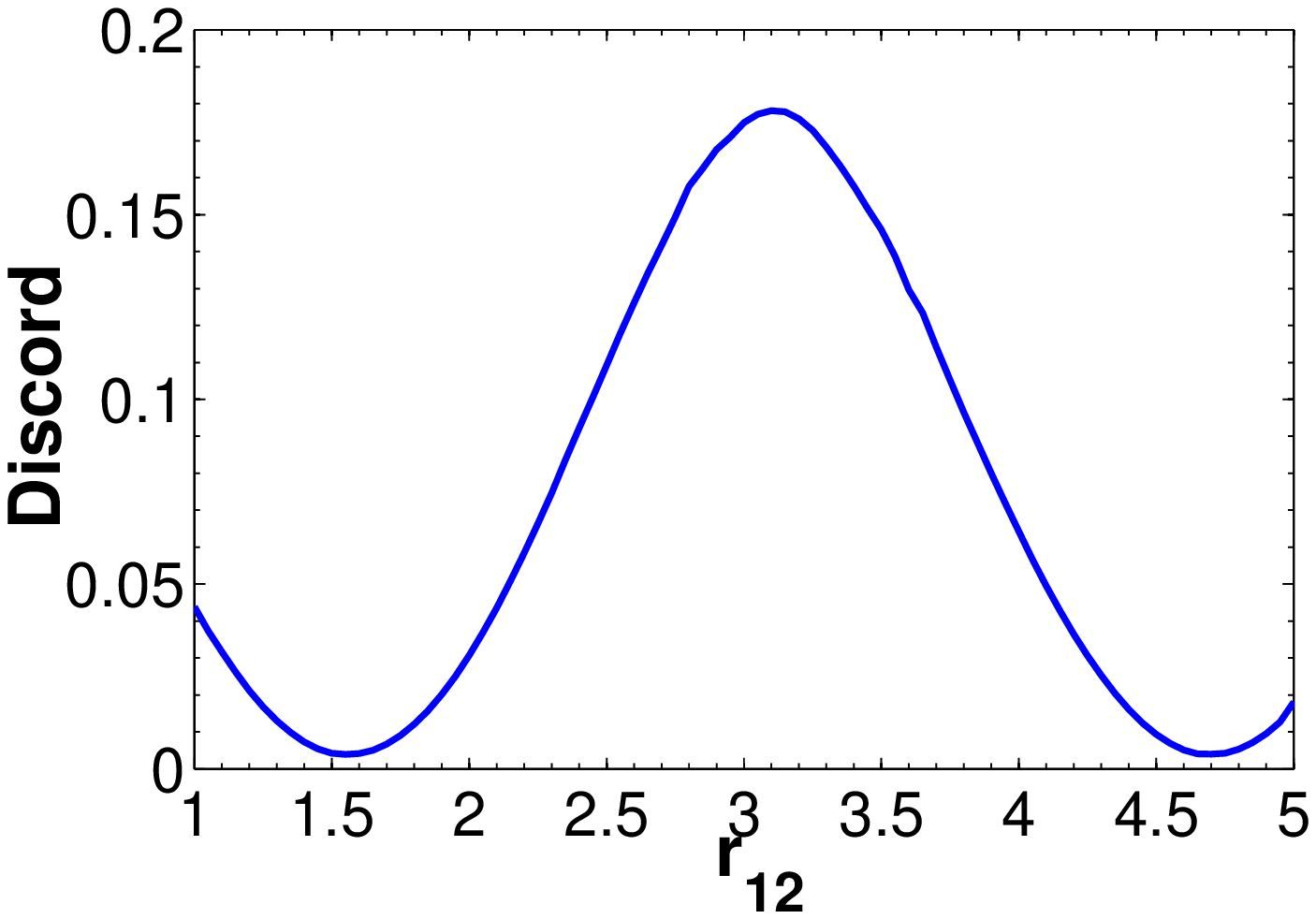}}\\
\caption{Figures (a), (b), (c) and (d) represent the evolution,  generated by a QND type of 
interaction, 
of concurrence, maximum teleportation fidelity $F_{max}$, test of Bell's inequality $M(\rho)$,
discord as a function of inter-qubit distance $r_{12}$.  Here 
temperature  $T = 0$, evolution time $t$ is 1.1 and 
bath squeezing parameter $r = -1$.}\label{fig3}
\end{figure}
\end{center}


\begin{figure}
\subfloat[]{\includegraphics[height=4.0cm,width=7cm]{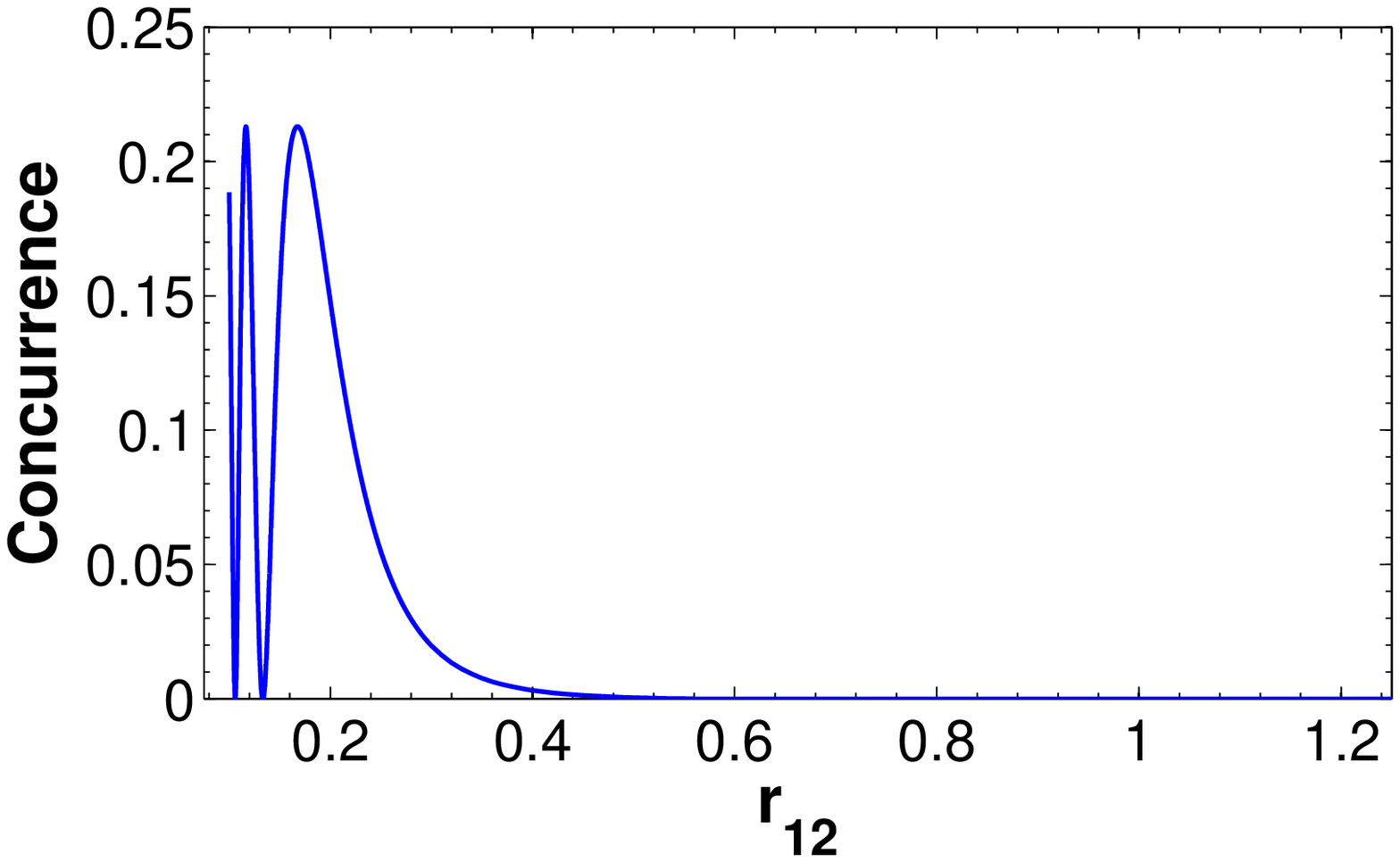}}
\subfloat[]{\includegraphics[height=4.0cm,width=7cm]{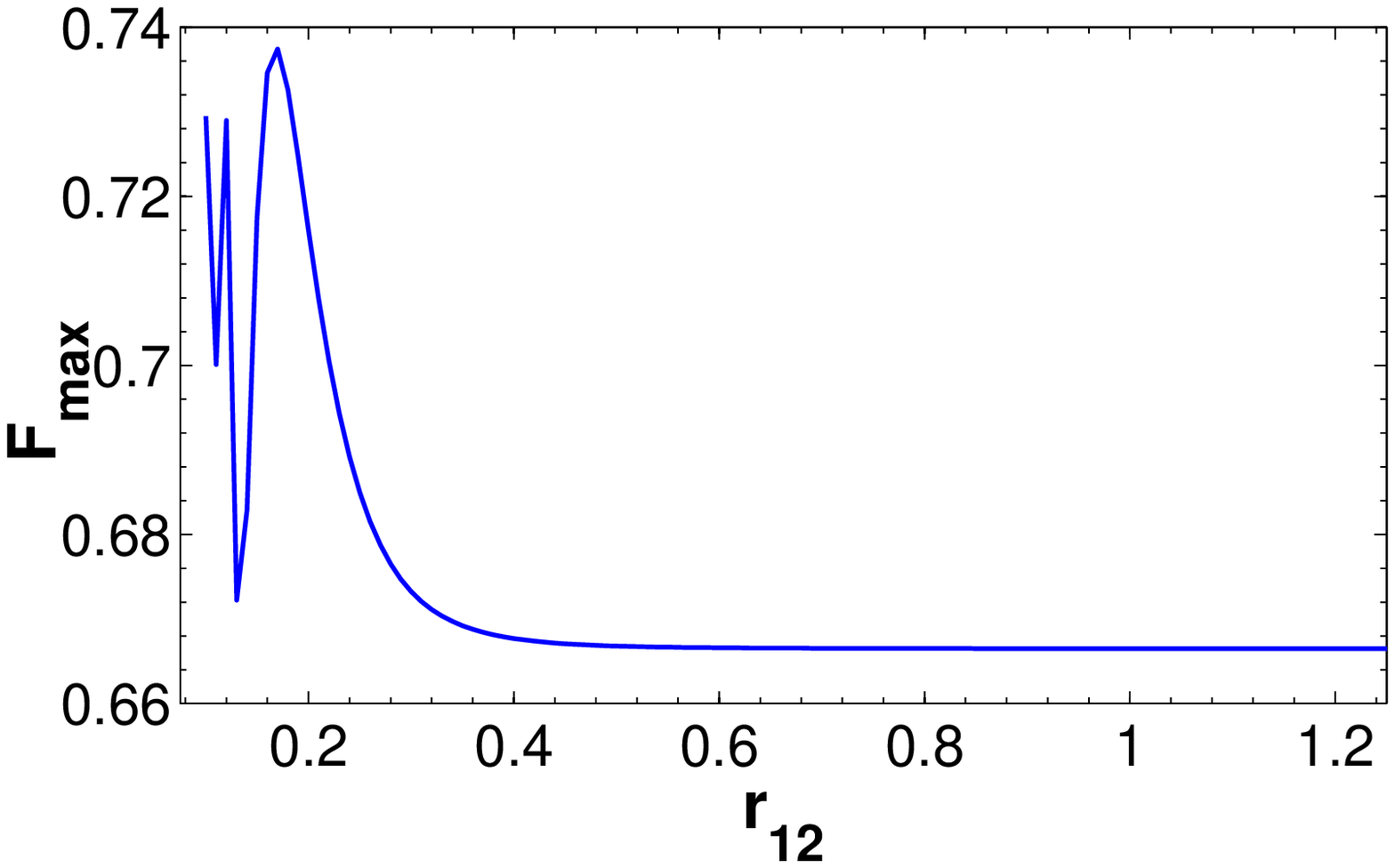}}\\
\subfloat[]{\includegraphics[height=4.0cm,width=7cm]{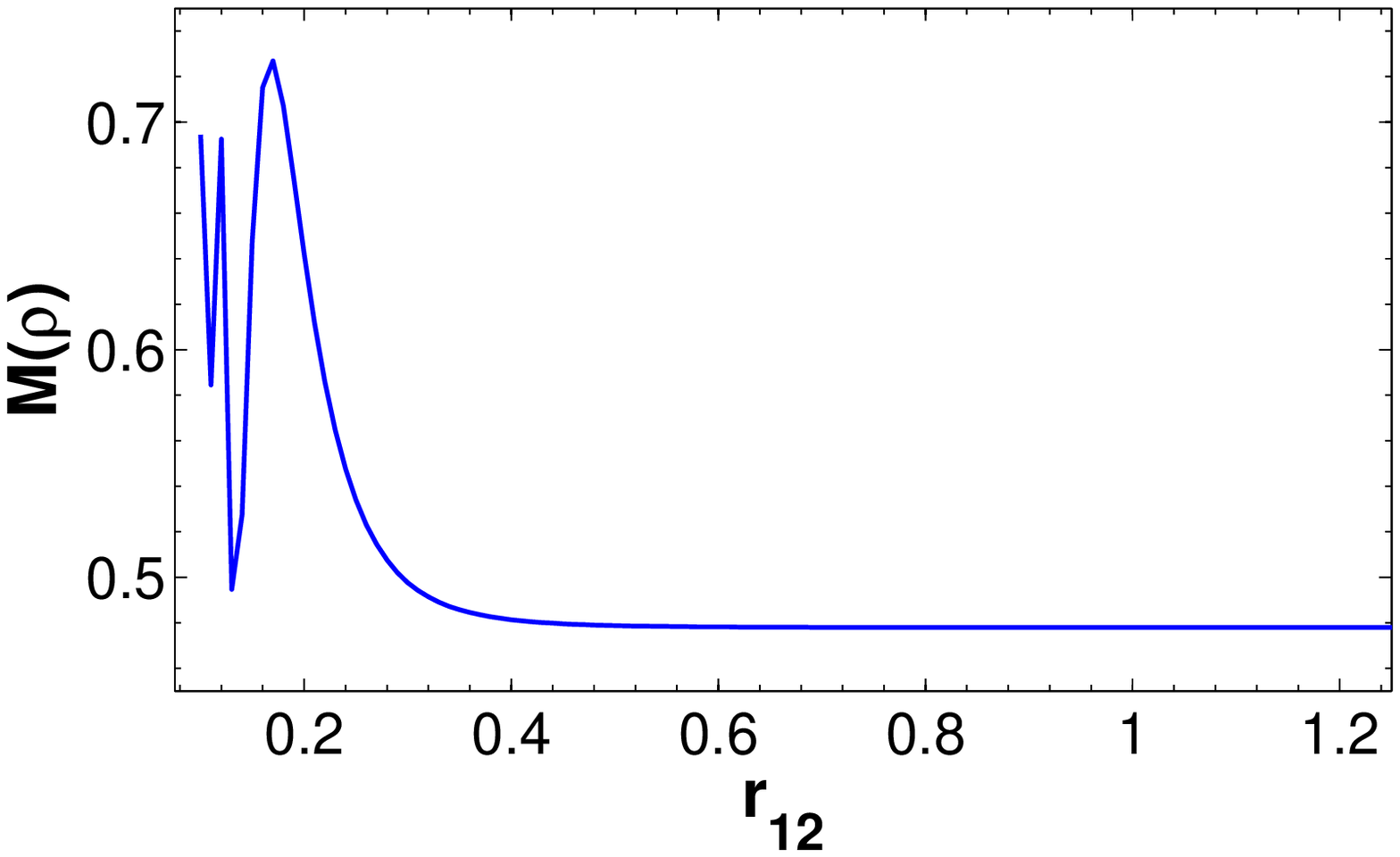}}
\subfloat[]{\includegraphics[height=4.0cm,width=7.0cm]{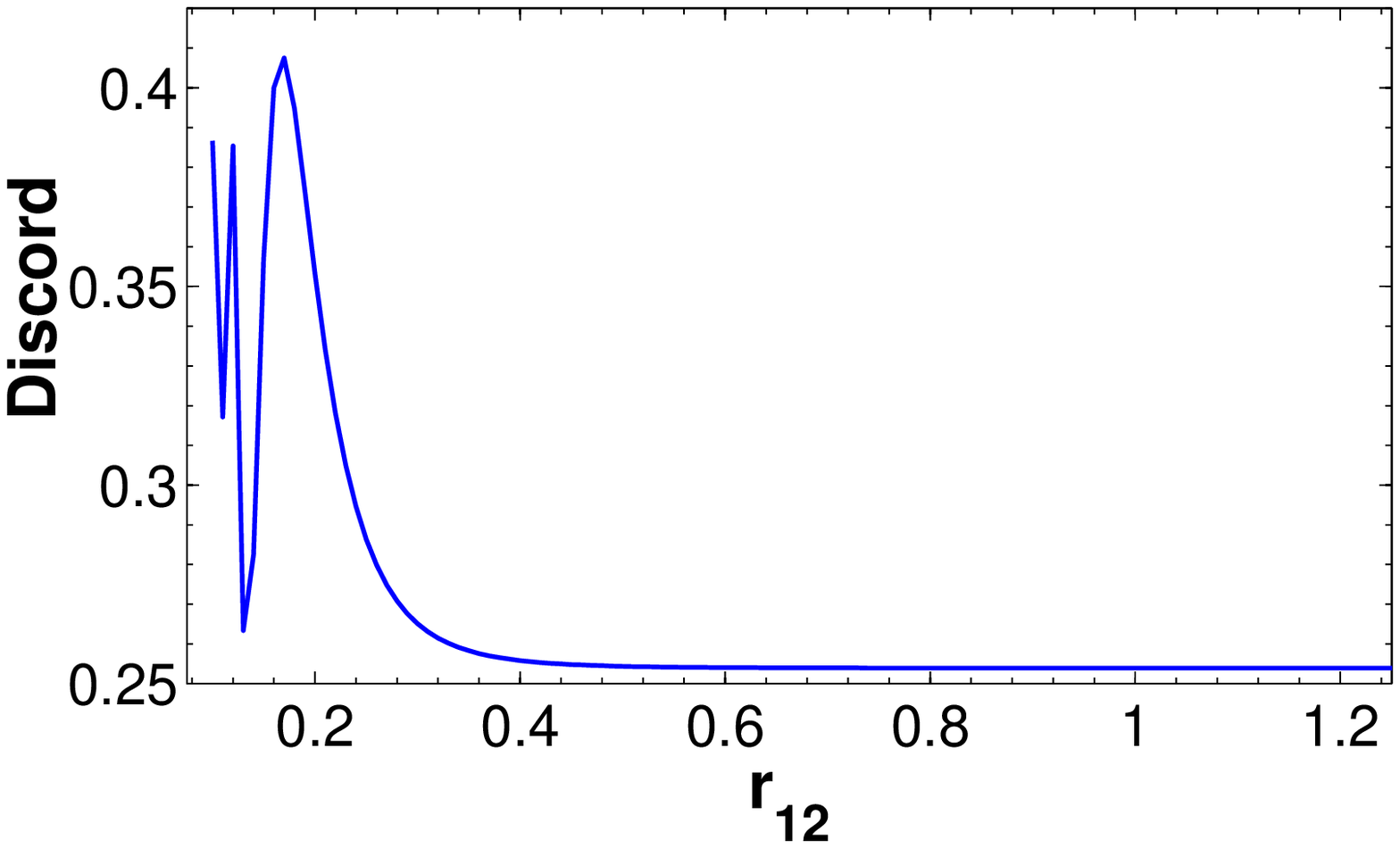}}\\
\caption{Quantum correlations in a two-qubit system undergoing a dissipative
evolution.  The Figs. (a), (b), (c) and (d) represent the evolution of
concurrence, maximum teleportation fidelity $F_{max}$, test of Bell's inequality $M(\rho)$,
discord as a function of inter-qubit distance $r_{12}$.  Here 
temperature  $T = 300$, evolution time $t$ is 0.1 and bath squeezing 
parameter $r =  -1$.}\label{fig4}
\end{figure}

\begin{figure}
\subfloat[]{\includegraphics[height=4cm,width=7cm]{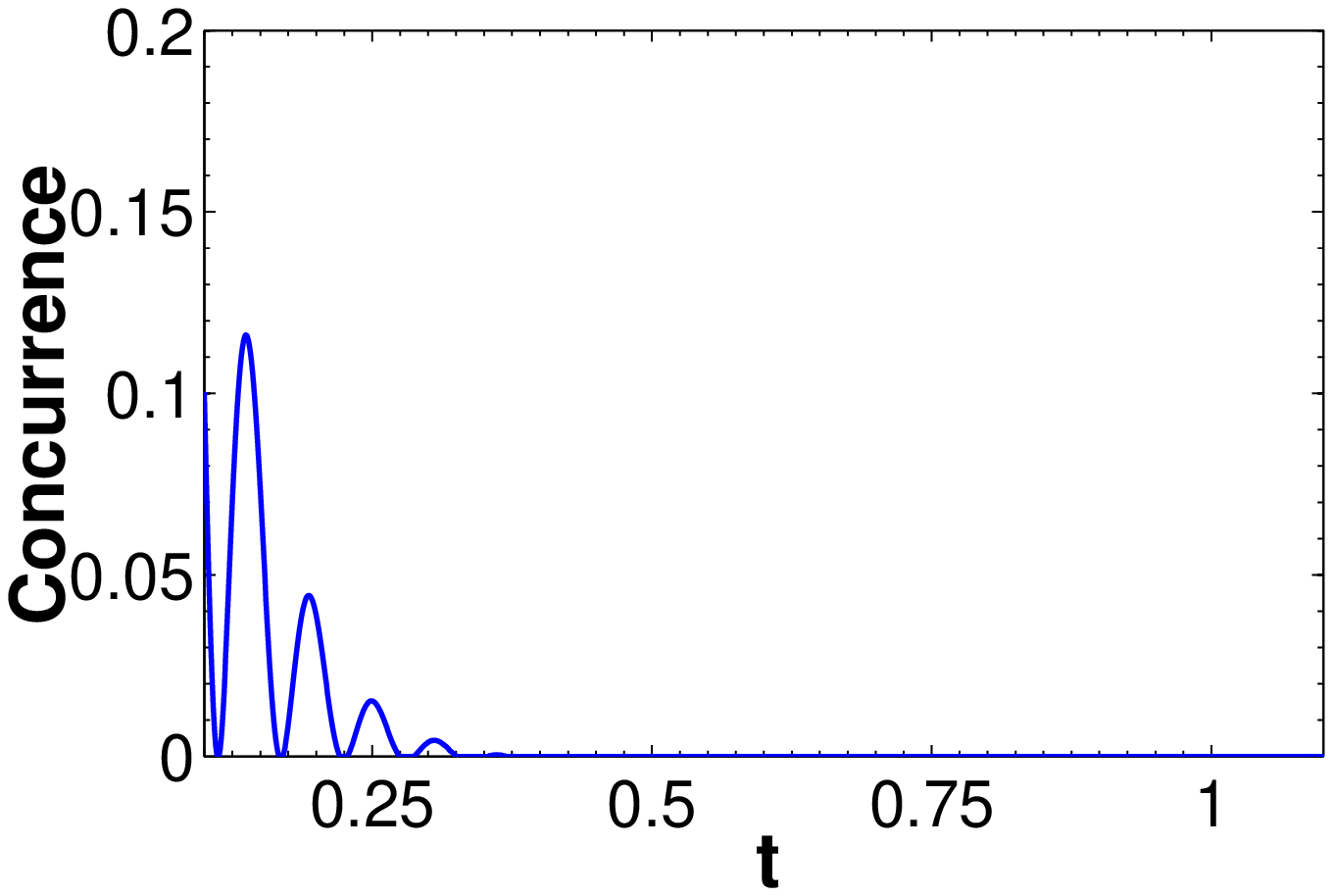}}
\subfloat[]{\includegraphics[height=4cm,width=7cm]{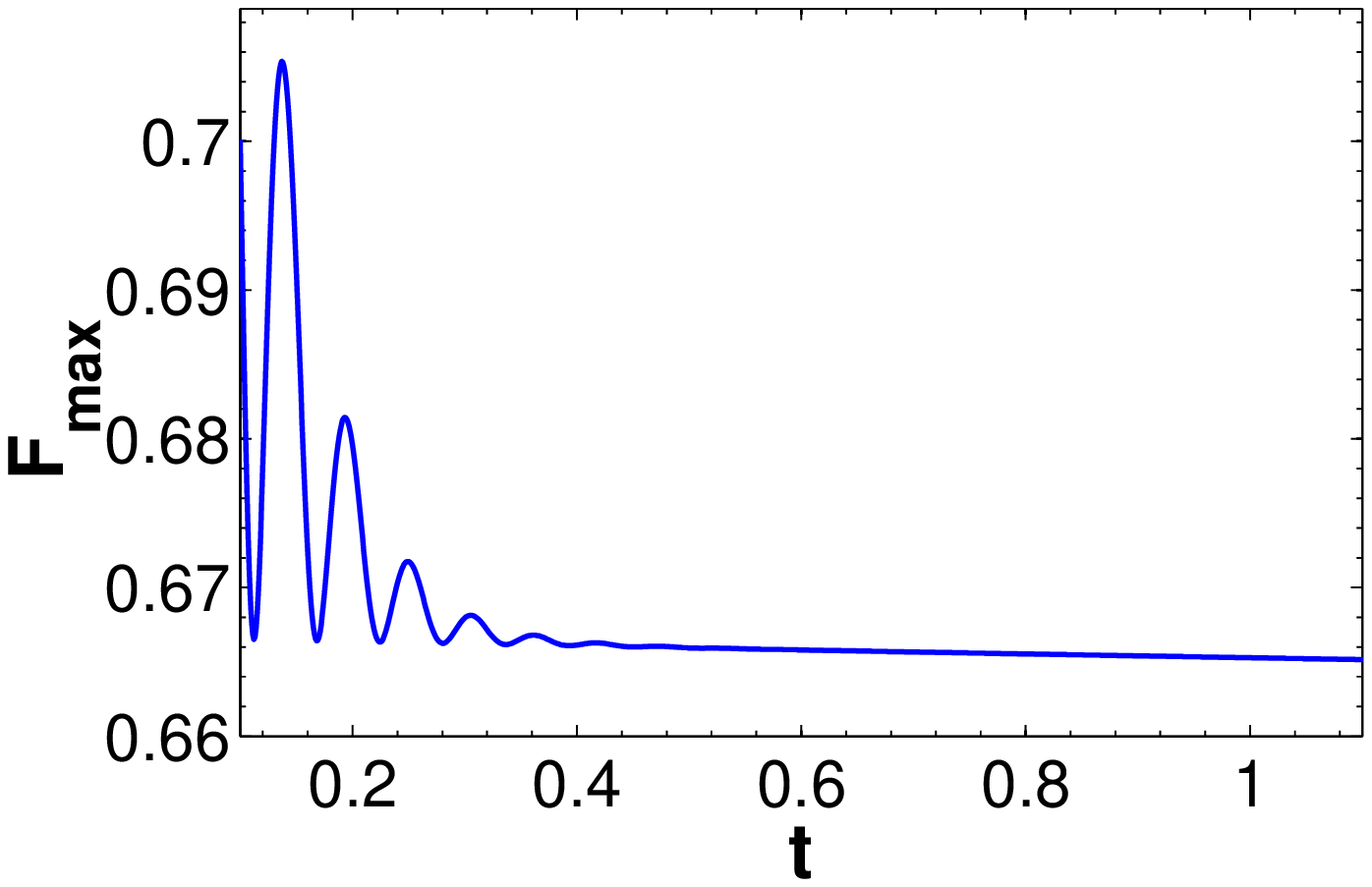}}\\
\subfloat[]{\includegraphics[height=4cm,width=7cm]{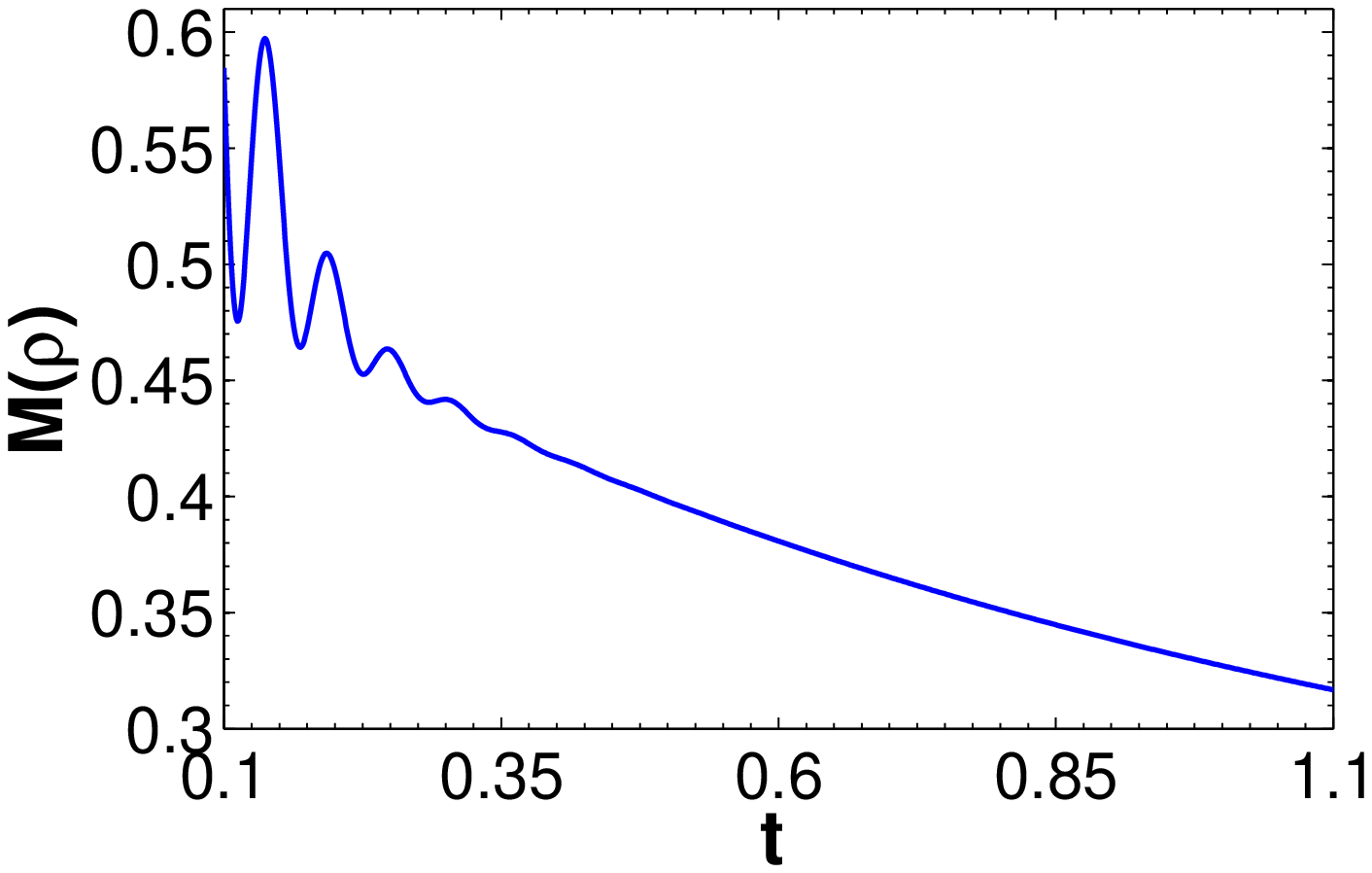}}
\subfloat[]{\includegraphics[height=4cm,width=7cm]{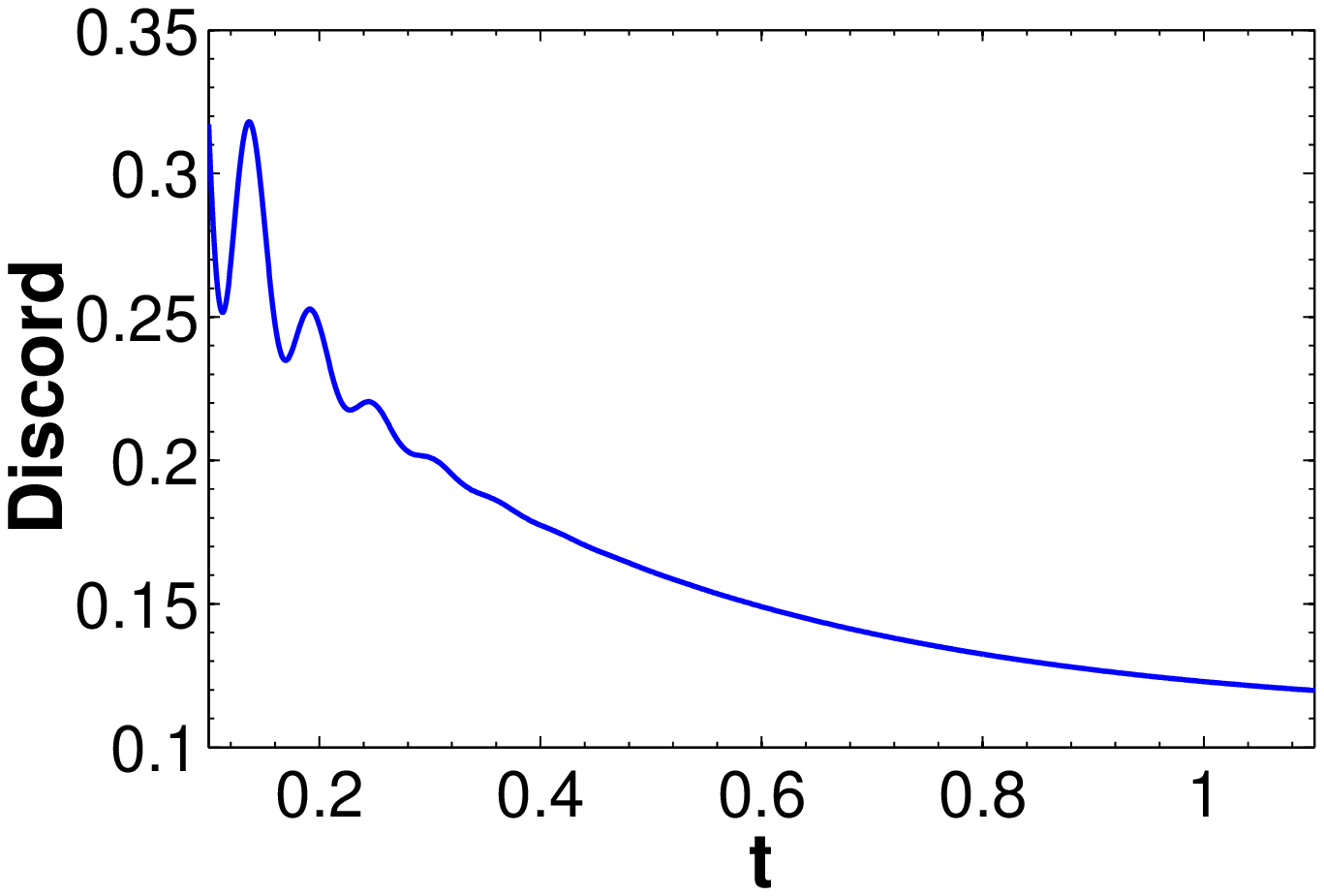}}\\
\caption{Figures (a), (b), (c) and (d) represent the evolution of
concurrence, maximum teleportation fidelity $F_{max}$, test of Bell's inequality $M(\rho)$,
discord with respect to the time of evolution $t$, evolving under a
dissipative interaction.  Here 
temperature  $T = 300$,  inter-qubit distance $r_{12} = 0.11$ and bath squeezing 
parameter $r =  -1$.}\label{fig5}
\end{figure}

\begin{center}
\begin{figure}
\subfloat[]{\includegraphics[height=4.0cm,width=7cm]{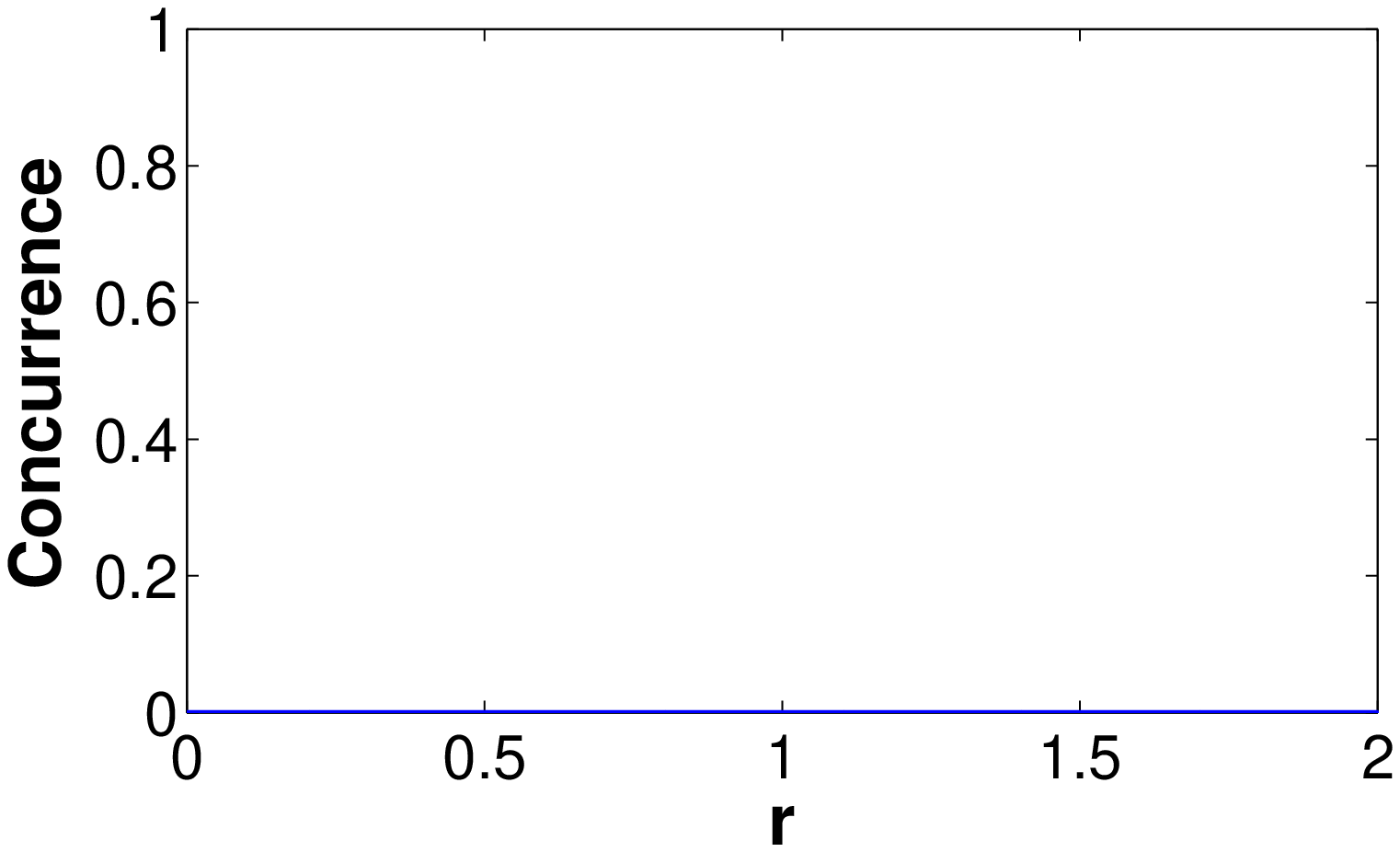}}
\subfloat[]{\includegraphics[height=4.0cm,width=7cm]{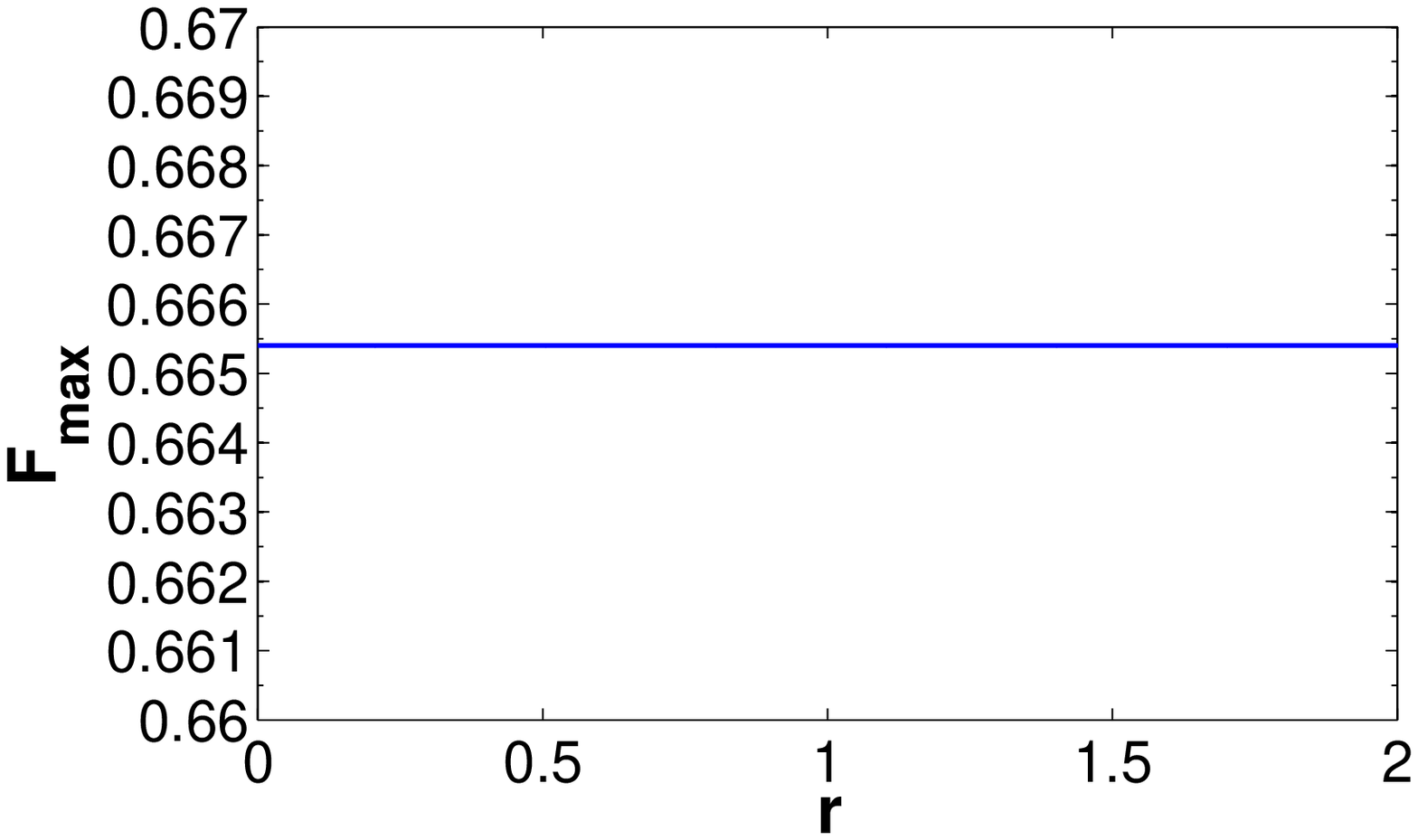}}\\
\subfloat[]{\includegraphics[height=4.0cm,width=7cm]{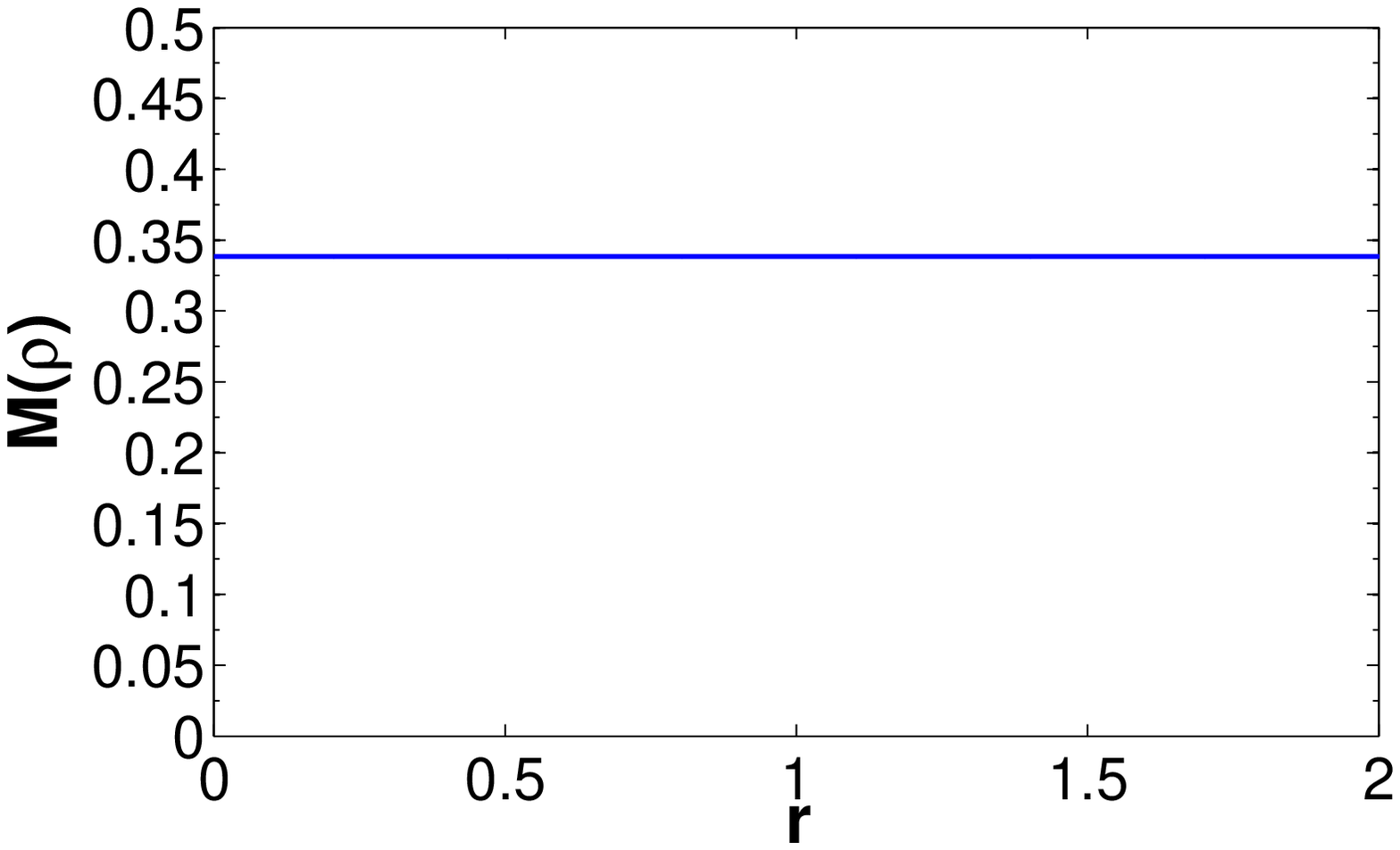}}
\subfloat[]{\includegraphics[height=4.0cm,width=7cm]{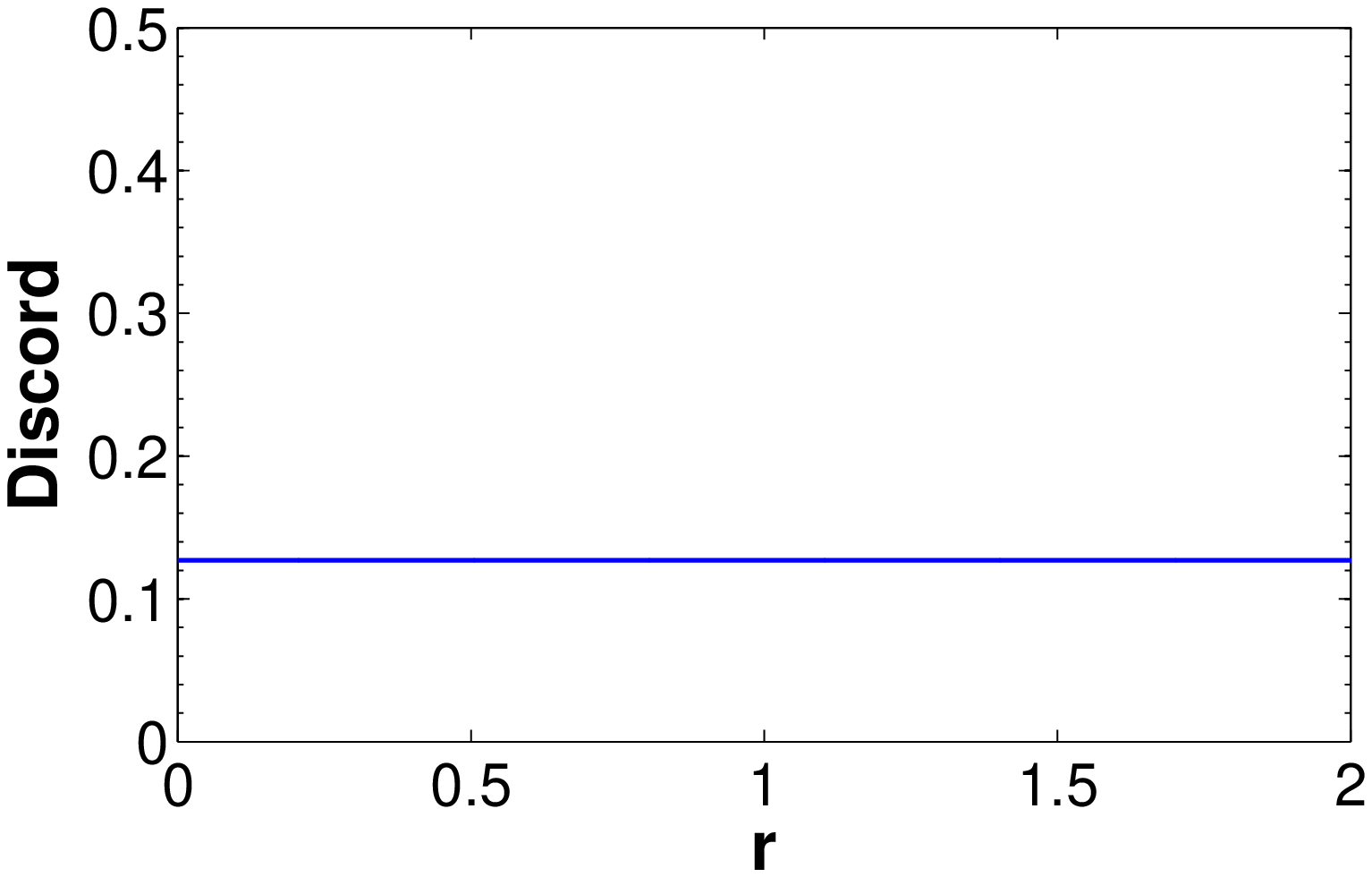}}\\
\caption{An example showing vanishing entanglement, but non vanishing discord, for a
dissipative two-qubit evolution. 
Figures (a), (b), (c) and (d) represent the evolution of
concurrence, maximum teleportation fidelity $F_{max}$, test of Bell's inequality $M(\rho)$,
discord with respect to the bath squeezing 
parameter $r $ ranging from $0$ to $2$.  Here 
temperature  $T = 10$, $t = 0.9$ and  inter-qubit distance $r_{12} = 0.09$.}\label{fig6}
\end{figure}
\end{center}

\begin{center}
\begin{figure}
\subfloat[]{\includegraphics[height=6.0cm,width=8cm]{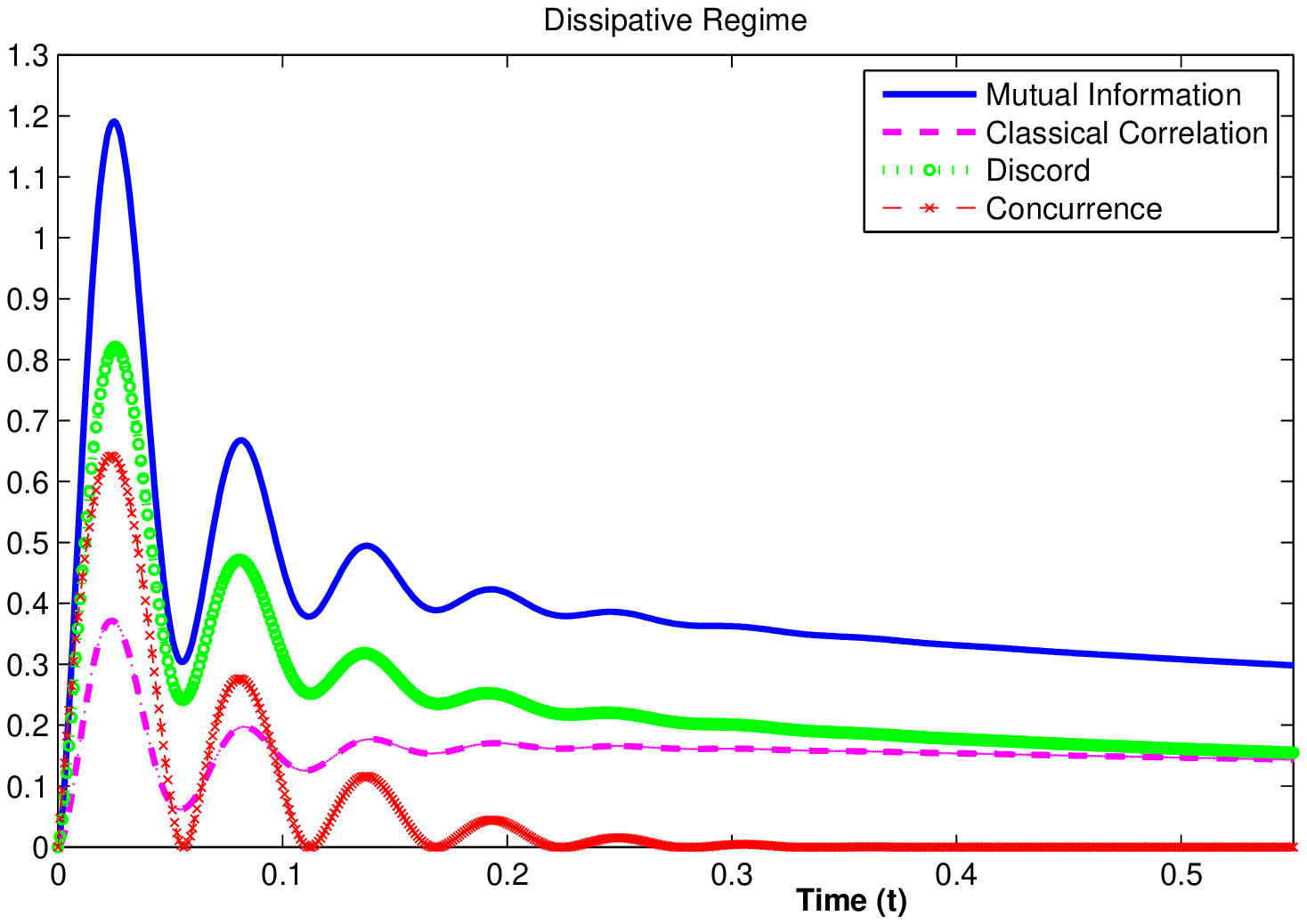}}
\subfloat[]{\includegraphics[height=6.0cm,width=8cm]{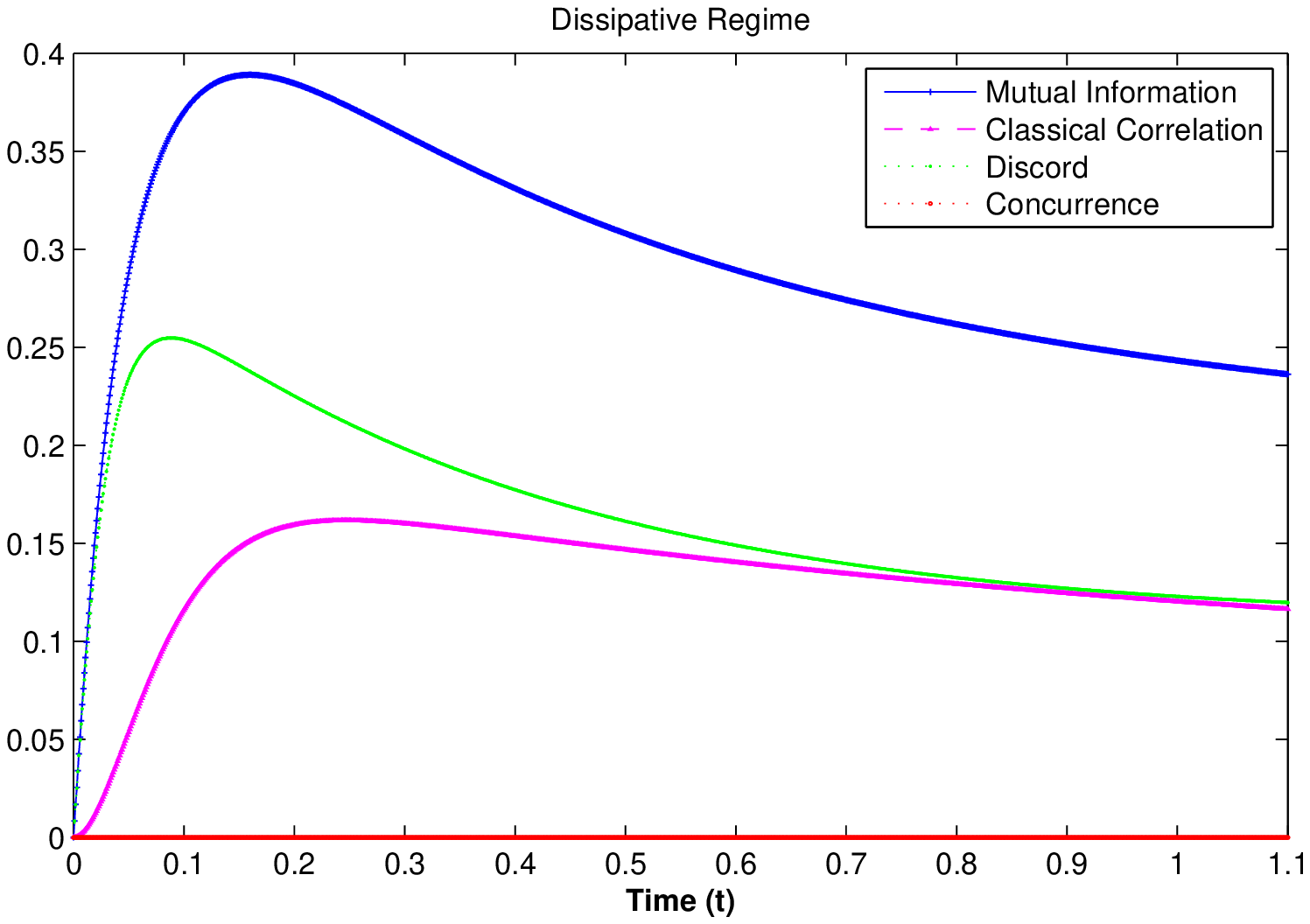}}\\
\caption{An example showing vanishing entanglement, but non vanishing discord, for a
dissipative two-qubit evolution. 
Figures (a)and (b) represent the evolution of mutual information (blue), quantum discord (quantum correlation) (green), concurrence (red) and classical information (pink)
with respect to the time of evolution $t$, evolving under a
dissipative interaction for collective and independent models, respectively.  Here  for (a)
temperature  $T = 10$, $r = 0$ and  inter-qubit distance $r_{12} = 0.11$ and for (b)
temperature  $T = 10$, $r = 0$ and  inter-qubit distance $r_{12} = 1.5$. }\label{fig7}
\end{figure}
\end{center}
\end{widetext}
\end{document}